\documentclass[%
reprint, superscriptaddress,
 amsmath,amssymb,
 aps,pre,
 nofootinbib
]{revtex4-2}
\usepackage{graphicx}
\usepackage{dcolumn}
\usepackage{bm}
\usepackage{dsfont}
\usepackage{amsmath}
\usepackage{todonotes}
\usepackage{siunitx}
\usepackage[normalem]{ulem}
\usepackage{float}
\usepackage{placeins}
\usepackage[utf8]{inputenc}
\usepackage{nicefrac}
\usepackage{xcolor}

\definecolor{linkColor}{rgb}{0,0.3,0.7}
\usepackage[colorlinks=true,
            allcolors=linkColor,
            pdfborder={0 0 0},
            pdfencoding = auto
            ]{hyperref}
            
\usepackage{url}

\begin{document}

\title{Phase Separation on Deformable Membranes: {\\} interplay of mechanical coupling and dynamic surface geometry}

\author{Antonia Winter}
\thanks{These authors contributed equally to this work}
\affiliation{Arnold Sommerfeld Center for Theoretical Physics and Center for NanoSciences, Ludwig-Maximilians-Universität München, Theresienstraße 37, 80333 Munich, Germany.}
\author{Yuhao Liu}
\thanks{These authors contributed equally to this work}
\affiliation{Arnold Sommerfeld Center for Theoretical Physics and Center for NanoSciences, Ludwig-Maximilians-Universität München, Theresienstraße 37, 80333 Munich, Germany.}
\author{Alexander Ziepke}
\thanks{These authors contributed equally to this work}
\affiliation{Arnold Sommerfeld Center for Theoretical Physics and Center for NanoSciences, Ludwig-Maximilians-Universität München, Theresienstraße 37, 80333 Munich, Germany.}
\author{George Dadunashvili}
\affiliation{Arnold Sommerfeld Center for Theoretical Physics and Center for NanoSciences, Ludwig-Maximilians-Universität München, Theresienstraße 37, 80333 Munich, Germany.}
\author{Erwin Frey}
\email{frey@lmu.de}
\affiliation{Arnold Sommerfeld Center for Theoretical Physics and Center for NanoSciences, Ludwig-Maximilians-Universität München, Theresienstraße 37, 80333 Munich, Germany.}
\affiliation{Max Planck School Matter to Life, Hofgartenstra{\ss}e 8, 80539 Munich, Germany.}

\date{\today}

\begin{abstract}
The self-organization of proteins into enriched compartments and the formation of complex patterns are crucial processes for life on the cellular level.
Liquid-liquid phase separation is one mechanism for forming such enriched compartments. 
When phase-separating proteins are membrane-bound and locally disturb it, the mechanical response of the membrane mediates interactions between these proteins.
How these membrane-mediated interactions influence the steady state of the protein density distribution is thus an important question to investigate in order to understand the rich diversity of protein and membrane-shape patterns present at the cellular level.
This work starts with a widely used model for membrane-bound phase-separating proteins. 
We numerically solve our system to map out its phase space and perform a careful, systematic expansion of the model equations to characterize the phase transitions through linear stability analysis and free energy arguments.
We observe that the membrane-mediated interactions, due to their long-range nature, are capable of qualitatively altering the equilibrium state of the proteins.
This leads to arrested coarsening and length-scale selection instead of simple demixing and complete coarsening.
In this study, we unambiguously show that long-range membrane-mediated interactions lead to pattern formation in a system that otherwise would not do so. 
This work provides a basis for further systematic study of membrane-bound pattern-forming systems.
\end{abstract}

\maketitle

\section{Introduction}
\label{sec:introduction}

Self-organization in the absence of external guiding cues is a key principle in the creation and maintenance of cellular structures. 
One important mechanism for forming structures and enriched compartments is liquid-liquid phase separation \cite{Brangwynne2009Germline, Hyman:2014}, whereby local interactions between components like proteins or lipids induce demixing into a dense and a dilute phase \cite{Doi2013Soft}. 
Phase separation also occurs on cellular lipid membranes, where it leads to the formation of domains such as lipid rafts \cite{John2005Traveling,Veatch2003Separation} or enriched protein clusters \cite{Groves2007Bending,Alonso2010Phase}.
Such domains are crucial for cellular functions such as polarization \cite{Simons2000Lipid}, sensing \cite{Postma2004Sensitization,Shelby2023Membrane}, and membrane transport \cite{Ikonen2001Roles}.

In all these cases, the lipid bilayer membrane acts as an elastically bending manifold with embedded proteins.
This can lead to mutual feedback between the spatial organization of membrane-bound proteins and the mechanical properties of the membrane via interactions that are influenced by the geometry of the membrane surface. 
For instance, intracellular proteins can alter properties like the bending rigidity of lipid bilayer membranes \cite{Leibler:1986,Girad2005Passive} or may cause deformations due to their intrinsic curvature \cite{Farsad:2003, prevost2015irsp53, Goychuk:2019}.
A well-studied instance of phase-separating membrane-bound proteins that induce curvature involves BAR domains, which form high-density and dilute regions on the membrane \cite{prevost2015irsp53, simunovic2016how}. 
Lipid-demixing in multi-component membranes is another example of biological liquid-liquid phase separation, in which lipid-localization can affect local membrane curvature \cite{Heinrich:2010} as well as the local protein composition \cite{Shelby2023Membrane}. 
This highlights the importance of the dynamic interplay between phase-separation dynamics and mechanical deformations of surfaces.

Theoretical and experimental studies have explored how membrane curvature influences the demixing and positioning of lipids and membrane-embedded proteins in response to externally induced (static) membrane deformations~\cite{Parthasarathy2006Curvature,Heberle2011Phase,Vandin2016Curvature}. 
In particular, it has been reported that local curvature induced by micropipette aspiration induces curvature-dependent lipid sorting in biomembranes \cite{Callan-Jones2011Curvature}. Other experimental studies demonstrated phase separation and sorting of lipid rafts in giant unilamellar vesicles upon induction of varying curvatures \cite{Roux2005Role,Semrau2009Membrane,Idema2010Membrane}.

The crucial role of membrane curvature for pattern formation of curvature-sensing and curvature-inducing proteins and lipids has also inspired various theoretical studies on the dynamic interplay between membrane mechanics and liquid-liquid phase separation \cite{Seifert1993Curvature,Lavrentovich2016First,Agudo-Canalejo2017Pattern,Magi2017Modelling}.
These theoretical analyses primarily focus on the onset of phase separation and membrane deformations from the initial spatially uniform and flat state, but often do not investigate the subsequent (nonlinear) dynamics leading to a steady state.
Two recent studies have gone beyond the emergence of patterns and studied these nonlinear dynamics that result from the interplay between phase separation and mechanical deformations \cite{Mahapatra:2021,yu2023pattern}.
Using an expansion of their model equations that assumes weak membrane deformations, these authors find that phase separation is altered by the membrane-mediated interactions between proteins~\cite{Mahapatra:2021} and that membrane-mediated lipid-lipid interactions lead to a length-scale selection of emerging patterns~\cite{yu2023pattern}. 
However, these theoretical studies leave aside two crucial aspects of the dynamics.
Firstly, the analysis in Ref.~\cite{yu2023pattern} neglects the impact of temporal changes in membrane surface area on protein and lipid concentrations, thereby only approximately preserving mass-conservation
on the deformable surface for weak deformations.
Second, both studies disregard significant contributions from spatial changes in the membrane metric, that affect local interactions between the phase-separating components and the entropic contributions mediated by changes in membrane area.
Consequently, the question of how the dynamic interplay between membrane geometries and curvature-inducing phase-separating proteins affects protein dynamics remains unresolved.

\begin{figure}[!tb]
\centering
\includegraphics[width=\linewidth]{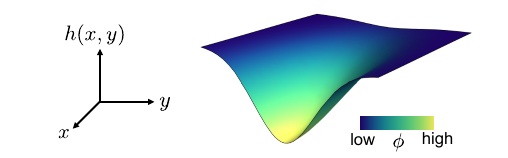}
\caption{
Illustration of the model for phase separation on a deformable membrane surface, in which protein density and membrane conformations are mechanically coupled.
On the membrane surface, represented in the Monge parametrization by a height field $h(x,y)$, proteins induce local curvature that scales with their density $\phi(x,y)$ (color code).
}
\label{fig:scheme}
\end{figure}

Here, we investigate how membrane-mediated interactions influence phase separation of membrane-bound proteins. 
We model the thermodynamics of the protein-lipid system using the Flory-Huggins free energy for a symmetric binary mixture~\cite{Flory:1941,Huggins:1941}.
The mechanical deformations of the liquid membrane are described by the Canham-Helfrich free energy~\cite{Canham:1970, Helfrich:1973}, which accounts for bending energy and surface tension. 
The dynamics of protein densities and membrane deformations are assumed to be coupled mechanically. 
Specifically, we follow the well-established literature \cite{Seifert1993Curvature} and examine proteins that cause spontaneous local curvature proportional to their density on the membrane.
This assumption is supported by experimental studies of BAR domains~\cite{Peter:2004, Saarikangas:2009, simunovic2016how,Heinrich2:2010} and clathrin complexes~\cite{Ford:2002, Kukulski:2012}, where the coupling forces the membrane to locally conform to the (potentially anisotropic) curvature induced by the protein coating, as illustrated in Fig.\@~\ref{fig:scheme}.
In a similar fashion, recent experiments emphasize the induction of membrane curvature by lipid-membrane-attached MinD proteins in giant vesicles which gives rise to blebbing and protein-controlled vesicle shape oscillations \cite{litschel2018beating,Christ2020Active}. Although the potential mechanism of curvature induction differs between these examples, spontaneous curvature models effectively capture the observed membrane deformations~\cite{Christ2020Active, Heinrich2:2010}.

What are the implications of the coupling between protein densities and membrane deformations for the dynamics of phase separation? 
While the proteins undergo phase separation and aggregation in droplet-like domains, they induce membrane deformations, which, in turn, alter the dynamics of the protein density.
In particular, a dynamically deforming membrane affects the protein dynamics through a spatially and temporally varying surface metric, changing the local concentration gradients of proteins and their interactions with neighboring proteins.

We present a generalized framework for understanding phase-separation dynamics on flexible, evolving membrane surfaces, such as bilayer membranes. 
Our approach is fully covariant, meaning it accounts for changes in the underlying geometry of the membrane as it deforms over time, independent of the chosen parametrization. 
The theory rigorously incorporates mass conservation and captures the mechanical feedback between phase separation of proteins and membrane curvature. Specifically, we account for the role of surface-bound proteins that mechanically couple to the membrane by inducing spontaneous curvature. 
These proteins locally bend the membrane by favoring regions of specific curvature, creating feedback mechanisms where phase separation and membrane deformations influence each other. Such interactions are particularly relevant in biological membranes, where curvature-generating proteins play a key role in organizing membrane domains and regulating cellular processes. 
This versatile framework can be extended to more complex systems, including multi-component phase separation and active, out-of-equilibrium processes that drive dynamic reorganizations of the membrane.

For the generic model system discussed, we find that mechanical coupling induces two distinct phase-separating (spinodal) instabilities: a classical long-wavelength Cahn-Hilliard instability, which is also present in the absence of geometric coupling, and a conserved Turing instability, characterized by a band of unstable modes that are bound away from zero wavevector but including it as a marginal mode~\cite{Frohoff2023Non}.
Depending on the mechanical coupling parameters, we observe steady-state patterns with geometrically arrested coarsening dynamics and we determine the pattern length scales based on thermodynamic arguments.
Overall, our study uncovers the mechanical coupling via deformable membranes as a mechanism behind membrane-mediated pattern formation and
provides a conceptual basis for systematic investigation of membrane-bound pattern-forming systems.

The paper is organized as follows:
Sec.~\ref{sec:model_description} introduces our model for liquid-liquid phase separation on dynamic membranes with mechanical coupling. We combine free energy contributions for the fluid membrane and protein-protein interactions and implement the geometric coupling via a spontaneous curvature of the proteins.
Additionally, we introduce dimensionless parameters and identify those crucial for tuning the strength of the interactions between the elastic (membrane) and chemical (protein) contributions to the free energy.
Finally, we give the dynamic equations that govern the evolution of the membrane shape and the protein density. These equations comprehensively describe the spatiotemporal variations of the membrane metric. 
They detail how the elastic properties of the membrane affect protein-protein interactions and mechanical coupling.
In Sec.~\ref{sec:coarsening}, we present the results of our numerical simulations, establishing that our system exhibits arrested coarsening and length-scale selection for strong coupling between membrane and protein densities.
In Sec.~\ref{sec:LSA} we connect numerical and analytical findings on the basis of linear stability analysis. We examine the problem from two perspectives. First, we conduct a classical linear stability analysis of the dynamic equations, identifying different types of instability with their associated dispersion relations of unstable modes. 
Second, we study the dispersion relation of an expanded, effective free energy of the proteins, which we determine by integrating out the degrees of freedom related to membrane fluctuations. 
Consistently, this approach provides analytical expressions for distinguishable instability types.
In Sec.~\ref{sec:lengt_scale_selection} we further develop this thermodynamic reasoning and successfully predict the onset of arrested coarsening. Additionally, we derive an analytic estimate for the mechanical length-scale selection of the final steady-state patterns.
Finally, we discuss our findings and shortly outline how our work may be generalized to other systems in Sec.~\ref{sec:discussion}.
This study is supported by six appendixes: Appendix~\ref{App:Derivation} provides the derivation of the dynamic equations, while Appendix~\ref{sec:numericalsim} gives details of the numerical simulations. In Appendixes~\ref{append:LSA} and \ref{append:LSA_FE}, we describe in detail the linear stability analysis based on the dynamic equations and the effective free energy, respectively. We comment on the comparison between the two approaches in Appendix~\ref{append:comparisonLSA}. Appendix~\ref{append:distmeasure} contains details on the distance measure applied to obtain the steady-state pattern length scales in the numerical simulations.

\section{Protein-membrane dynamics}
\label{sec:model_description}

To investigate the influence of mechanical interactions on the phase separation of proteins, we construct a minimal dynamical model that takes into account both protein-protein and membrane-mediated interactions. 
We assume that the protein-membrane system relaxes towards thermodynamic equilibrium via gradient dynamics that respect conservation laws. 
To model this process, we establish dynamic equations based on the principles of non-equilibrium thermodynamics.
Crucially, when formulating the free energy functional and the corresponding dynamic equations, one has to consider dynamical changes in the surface metric that arise from the deformations of the membrane surface.

\subsection{Free-energy functional}

We base our study on the free energy functional
\begin{equation} 
    \mathcal{F} [\phi,C] 
    = 
    \int \mathrm{d}A \, 
    \Big\{
    f [\phi, g_{ab}] + e [\phi,C] 
    \Big\}
    \,,
    \label{eqn:freeenergyfunc}
\end{equation}    
comprising of the functionals for the free energy densities of the proteins on the membrane, $f [\phi, g_{ab}]$, and the local conformations of the membrane itself, $e [\phi,C]$. 
Both of these functionals, as we will discuss in detail below, depend on the protein area fraction $\phi$ on the membrane and the membrane conformation, given in terms of the curvature $C$, which is the sum of the principal curvatures, and the metric tensor $g_{ab}$ of the membrane surface, where ${a,b \in \{1,2\}}$.
Moreover, for a parametrization $(u,v)$ of the membrane surface, the infinitesimal surface element is given by ${\mathrm{d} A = \sqrt{g} \, \mathrm{d}u \, \mathrm{d}v}$, where $g$ is the determinant of the metric tensor of the curved membrane surface.

We assume that the functional of the free energy density for the proteins is given by the standard Flory-Huggins (FH) theory for a symmetric binary mixture (of proteins and lipids)~\cite{Flory:1941, Huggins:1941, Desai:2009} 
\begin{align}
\label{eq:free_energy_density_FH}
    f [\phi, g_{ab}] 
    =& \, \rho_\text{s} \, k_\text{B}T
    \Big[ f (\phi) + \frac{\chi}{4\rho_\text{s}}\left| \nabla \phi \right|^2
    \Big]\,,
\end{align}
with the local part 
\begin{equation}
    f (\phi) 
    :=  \phi\ln\phi + (1-\phi)\ln(1-\phi)  + \chi \, \phi \, (1-\phi) 
    \,.
\end{equation}
In particular, $\phi$ denotes the protein area fraction on the membrane with a saturation density ${\rho_s = 1/v}$ given as the inverse of the two-dimensional molecular volume $v$.
In the following, we will measure all lengths in units of ${a\equiv \sqrt{v}}$, setting ${\rho_s=1}$.
While the first two terms describe entropic mixing, the terms proportional to the (dimensionless) Flory-Huggins parameter $\chi$ characterize the interaction between the proteins. 
Since the membrane is generally deformed, the differential operator $\nabla$ denotes the covariant derivative on the curved membrane, such that ${\left| \nabla \phi \right|^2 = (\partial_a \phi )(\partial_b \phi)g^{ab}}$. Thus, the FH free energy density becomes a functional of the membrane conformation.

For the functional of the free energy density of the membrane conformations, we assume a Canham-Helfrich form~\cite{Canham:1970,Helfrich:1973}
\begin{equation}
\label{eq:free_energy_density_CH}     
      e [\phi, C] 
      \equiv
      e (\phi, C) 
      = 
      \frac{\kappa}{2} 
      \big( 
      C  - C_0 \phi 
      \big)^2 
      + 
      \sigma 
      \,.
\end{equation}
It accounts for the free energy costs associated with membrane curvature $C$ and area changes, characterized by two parameters, the surface tension $\sigma$ and bending rigidity $\kappa$. Note that the microscopic surface tension of lipid bilayer membranes is intrinsically high \cite{Morris:2001}. The quantity $\sigma$ considered here represents an effective mesoscale surface tension that originates from thermal membrane fluctuations. These fluctuations modulate the membrane surface, storing excess area that can contribute to the membrane’s mechanical response \cite{Deserno:2014}.
We consider a total spontaneous membrane curvature mediated by the proteins bound to the membrane. 
For simplicity, we assume a linear dependence on the protein density $\phi$ with a proportionality factor $C_0$~\cite{Gozdz:2006}, which describes the mechanical coupling in the system. Hence, the coupling constant $C_0$ determines the extent to which the proteins influence the curvature of the membrane.

Assuming the absence of overhangs, we describe the membrane deformation using a Monge parametrization, representing the membrane surface as a height profile ${\mathbf{r} = (x,y,h(x,y))^T}$. 
Then, the membrane curvature \cite{Deserno:2014} is given by  
\begin{equation}
\label{eq:curvature_monge}
	C 
	= 
	\nabla_{\perp} \cdot 
	\bigg[   
    \frac{\nabla_{\perp} h}{ \sqrt{1+(\nabla_{\perp} h)^2}}
	\bigg] \, , 
\end{equation}
where ${\nabla_{\perp} = \hat{\mathbf{x}} \, \partial_x+\hat{\mathbf{y}} \, \partial_y}$ denotes the two-dimensional gradient operator on the base plane, with unit vectors $\hat{\mathbf{x}}$, $\hat{\mathbf{y}}$ spanning the parametrization domain. The metric tensor is given by 
\begin{equation}
\label{eq:metrictensor}
    g_{ab} = \begin{pmatrix}
1 + \left(\partial_x h\right)^2 & \left(\partial_x h \right)\left(\partial_y h \right) \\
\left(\partial_y h \right)\left(\partial_x h \right) & 1 + \left(\partial_y h\right)^2 
\end{pmatrix}
\end{equation}
and its determinant is ${g = 1 + (\nabla_{\perp} h)^2}$.

\subsection{Coupled gradient dynamics of protein density and membrane conformation}

We assume that the temporal evolution of the membrane surface ${\mathbf{r} = (x,y,h(x,y))^T}$ and the density field $\phi(x,y,t)$ follow gradient dynamics, neglecting hydrodynamic effects other than friction.
For the fluid membrane conformations, we use relaxational dynamics 
\begin{align}
\label{eqn:h_eq}
    \mathcal{D}_t \mathbf{r} 
    &= 
    - \gamma \,
    \frac{\delta \mathcal{F}}{\delta \mathbf{r}} \,,
\end{align}
where $\gamma$ is a positive Onsager coefficient; accounting for fluid dynamics beyond friction would require a more elaborate approach~\cite{Frey:1991}. The material derivative $\mathcal{D}_t \mathbf{r} = \partial_t \mathbf{r} - v_\mathrm{coord}^a \mathbf{t}_a$ serves as a link between the chosen Monge parametrization, where $\partial_t \mathbf{r} = (0,0,\partial_t h)^T$, and the actual movement of the membrane material points $\mathcal{D}_t \mathbf{r}$ \cite{Wuerthner:2023}. Here, $\mathbf{t}_a$ represents the tangent vector, and the introduced coordinate velocity $v^a_\mathrm{coord}$ is essential for ensuring that the density is correctly advected along the material points of the membrane rather than along the chosen coordinate axes.

To set up the equation for the protein dynamics, one starts from the conservation law for the number of proteins, which for an arbitrary domain $\Omega(t)$ with boundary$\partial \Omega(t)$ on the membrane surface reads
\begin{align}
   \frac{\mathrm{d}}{\mathrm{d}t} 
    \int_{\Omega(t)} \text{d}A \ \phi 
    =
    -\int_{\partial \Omega(t)} \text{d}l \ \mathbf{J} \cdot \mathbf{n} \, ,
\end{align}
where $\mathbf{J}$ denotes the protein density current, $\mathbf{n}$ the outer normal of $\Omega(t)$ and ${\text{d}A = \sqrt{g} \,  \text{d}x \text{d}y}$ the surface element. Since generally the deformation direction does not align with the chosen coordinate system, the boundary of the domain can be time dependent.
Using  Stoke's theorem, Reynolds transport theorem
and noting that a dynamically deforming membrane surface has a time-dependent metric, this results in a covariant continuity equation \cite{Salbreux2017Mechanics,Wuerthner:2023}
\begin{equation}
      \frac{1}{\sqrt{g}} \, 
      \partial_t \left(\sqrt{g} \, \phi \right) = - \nabla_a \left( J^a - v_\mathrm{coord}^a \phi \right)
      \, . 
\label{continuitytimedependentmetric}
\end{equation} 
Note, that the term $\nabla_a \, v_\mathrm{coord}^a \phi$ is a consequence of the chosen coordinate system and not a physical advection term.
For systems relaxing to thermodynamic equilibrium the particle current $\mathbf{J}$ is given by the gradient of the chemical potential, ${\mathbf{J} = - M \nabla \mu}$, with a constant mobility $M$ related to the diffusion constant $D$ and the thermal energy scale by the Einstein relation ${M = D/k_{\text{B}}T}$~\cite{degroot:2013}.
The chemical potential $\mu$ is given by the functional derivative of the free energy functional with respect to the area fraction, ${\mu = \delta {\cal F} / \delta \phi}$. 
This results in a covariant form of the Cahn-Hilliard equation~\cite{Cahn:1958, Cahn1961spinodal} that accounts for a time-dependent metric, 
\begin{align}
    \frac{1}{\sqrt{g}} \, 
    \partial_t
    \left(
        \sqrt{g} \phi
    \right)    
    = \nabla_a \left(
    M \,\nabla^a\frac{\delta \mathcal{F}}{\delta \phi} + v_\mathrm{coord}^a \phi\right)
    \, ,
\label{eqn:phi_eq}
\end{align}
where $\nabla_a$ describes the covariant derivative 
\begin{equation}
    \nabla_a \,v_\mathrm{coord}^a = \frac{\partial_a \left( \sqrt{g} \,v_\mathrm{coord}^a \right)}{\sqrt{g}}
\end{equation}
and $\nabla^2$ is the Laplace-Beltrami operator
\begin{equation}
    \nabla_a \nabla^a 
    = 
    \frac{1}{\sqrt{g}} \, 
    \partial_a
    \left[
    \sqrt{g} \, g^{ab}\partial_b 
    \right] \, .
\end{equation}
Note that Eq.~\eqref{eqn:phi_eq} gives the protein dynamics in a covariant form and therefore, is valid for arbitrary surfaces that give rise to the metric $g_{ab}$. Thus, it introduces no further approximations beyond the Monge representation. It should therefore be exact as long as overhangs in the membrane conformations are negligible.

Taken together, Eqs.~\eqref{eqn:h_eq} and \eqref{eqn:phi_eq} describe the coupled close-to-equilibrium dynamics of proteins and fluid membranes. 
The proteins undergo phase separation, simultaneously inducing spontaneous curvature on the deformable membrane. This mechanical coupling influences the bending dynamics of the membrane, which in turn affects the protein dynamics through geometric effects encoded in the dynamic metric of the membrane surface.

\subsection{Nondimensionalization and choice of parameters}
\label{subsec:nondimensionalization}

We now further non dimensionalize the set of dynamic equations~\eqref{eqn:h_eq} and \eqref{eqn:phi_eq}.
In addition to rescaling length in units of the protein size ${a \equiv 1/\sqrt{\rho_\text{s}}}$  we rescale time by the corresponding diffusive timescale ${\tau \equiv 1/(D \rho_\text{s})}$.
The main dimensionless parameters determining the mechanical feedback are the bending rigidity ${\kappa/k_\text{B}T\rightarrow\kappa}$, expressed in units of the thermal energy $k_\text{B}T$, the FH parameter $\chi$, and the protein-induced curvature ${ C_0 / \sqrt{\rho_\text{s}}\rightarrow C_0}$ in units of the protein size.
To focus on the impact of these key parameters on the mechanical feedback, we keep the dimensionless surface tension ${ \sigma / (k_\text{B}T\rho_\text{s})\rightarrow \sigma}$ and the Onsager coefficient ${ \gamma \rho_\text{s} k_\text{B}T / D \rightarrow \gamma}$ constant throughout the following analysis. 
 
All parameter ranges are chosen to mirror typical values found for membrane proteins, for instance, BAR domains  or clathrin~\cite{prevost2015irsp53, Bucher:2018};
see Table \ref{table:parameters}.
Accordingly, we choose the system size ${L = 2\, \si{\micro \meter}}$, the bending rigidity in the range ${\kappa = 0-40\, k_\text{B} T}$ \cite{Fowler:2016}, the spontaneous protein curvature in the range ${C_0 = 0-0.04 \, \si{\per \nano \meter}}$ \cite{Quemeneur:2014}, the membrane surface tension  ${\sigma \approx 1\cdot 10^{-4} \, \si{\pico\newton \per \nano \meter}}$ \cite{Baumgart:2003}, and the protein size ${1 / \sqrt{\rho_\text{s}} \approx 55 \, \si{\nano \meter}}$ \cite{Milo:2016}. 
Note that there is a wide range of values of $\sigma$ reported in the literature since it can describe different physics. 
Here we choose a small value of sigma, which represents the cost of pulling area from a reservoir of thermal fluctuations.

\begin{table}[tb!]
\centering
\caption{Ranges of values of the dimensionless parameters in units of the chosen scales (in brackets) taken from Refs.~\cite{Fowler:2016,Quemeneur:2014, Baumgart:2003, Milo:2016}, as specified in the paper. 
}
\medskip
\begin{tabular}{p {7em} l l l }
\hline 
$\chi$ 
& $>0 $ & \phantom{abc} Protein interaction \\ [1.1mm]
$C_0 \, [\sqrt{\rho_s}]$ & $0 - 1.2$ & \phantom{abc} Protein curvature\\[1.1mm]
$\sigma \, \left[k_\text{B}T \rho_\text{s}\right]$ & $0.16$ & \phantom{abc} Surface tension \\[1.1mm]
$\kappa \, \left[k_\text{B}T\right]$ & $0-40$  & \phantom{abc} Bending rigidity \\[1.1mm]
$\gamma \, \left[D /k_\text{B}T \rho_\text{s}\right]$ & 1 & \phantom{abc} Onsager coefficient \\
\hline
\end{tabular}
\label{table:parameters}
\end{table}

To focus on the impact of membrane-mediated interactions on the evolving patterns, we fix the protein interaction parameter to ${\chi = 3}$ throughout this study, unless stated otherwise.
Due to mass conservation, the total protein mass is fixed and is determined by the initial condition.  
As our initial condition, we choose a flat membrane with a homogeneous density of ${\phi_0(x, y) = 0.3}$, thus the initial average area fraction is ${\bar{\phi} = \frac{1}{A}\int\mathrm{d}A \, \phi(x,y) = 0.3}$.\footnote{Note that in our system, the total mass $\int\mathrm{d}A\,\phi(x,y)$ is conserved. Thus, if the considered surface area $A=\int\mathrm{d}A$ changes over time due to deformations, this will result in dynamic changes in the average protein area fraction $\bar{\phi}$.} 
This choice results in spinodal decomposition on flat membranes, as for ${\bar \phi = 0.3}$ the FH parameter ${ \chi =3}$ significantly exceeds the spinodal line ${ \chi_s \approx 2.38}$.
Additionally, we assume that the timescale of the protein dynamics is comparable to that of the membrane dynamics ${\gamma = 1}$.

\subsection{Dynamic equations and their interpretation}

We close this section by summarizing the dynamic equations in their dimensionless form.
The dynamics of the protein density is given by a generalized  Cahn-Hilliard equation that fully accounts for the dynamics of the metric tensor,
\begin{align}\label{eq:dyn_phi_nd}
    \frac{\partial_t (\sqrt{g} \,  \phi)}{\sqrt{g}} 
    =& 
    \nabla_a 
    \Big[ \nabla^a \left(
    \partial_\phi f(\phi) +  \partial_\phi e (\phi,C)
    -
    \frac{\chi}{2} \nabla^2 \phi
    \right) \nonumber \\
    & +  v_\mathrm{coord}^a \phi \Big] \, ,
\end{align}
where the additional term ${\partial_\phi e (\phi,C) = - \kappa C_0 \, (C - C_0 \phi)}$ describes the coupling of the protein density to the membrane curvature. 
The local term now reflects the competition between minimizing the free energy of the protein system and minimizing the mechanical deformation energy of the membrane.

In the absence of a coupling to a protein density, the membrane dynamics is driven by the surface tension $\sigma$ and the bending rigidity $\kappa$ of the membrane \cite{ou-yang1989Bending}
\begin{align}
    \frac{1}{\sqrt{g}\,  \gamma} \,
    \partial_t h 
    &= \sigma \, C 
    - 
    \kappa \, \left[\nabla^2 C +\, \frac{C}{2} \left(C^2 - 4 C_\mathrm{G} \right)\right]
    \, , 
\label{eq:shape_equation_membrane}
\end{align} 
where ${C_\text{G} = g^{-2} \det[\partial_a \partial_b h]}$ denotes the Gaussian curvature. 
Accounting for the mechanical coupling
we obtain (see Appendix~\ref{App:Derivation} for details) 
\begin{align}\label{eq:dyn_h_nd}
    \frac{1}{\sqrt{g}\,\gamma}
    \, \partial_t h 
    = \,
    & 
    \big[ 
    f [\phi, g_{ab}] + e(\phi,C)
    \big] \, C 
    - \kappa \nabla^2 (C - C_0 \phi) 
    \notag \\ 
    &+
    \kappa \, (C - C_0 \phi) \, (-C^2 + 2 C_\text{G})
    \notag \\
    &- 
    \tfrac{1}{2} 
    \chi \, 
    C^{ab} \, (\partial_a \phi) (\partial_b \phi) \, ,
\end{align}
where ${C_{ab} = g^{-1/2} \partial_a \partial_b \, h}$ denotes the curvature tensor.
This still has the overall structure of the dynamic shape equation, Eq.~\eqref{eq:shape_equation_membrane}, with a few essential new features. The first term now has as a prefactor the total free energy functional including the contributions from the protein interactions. 
The spatial gradients in the curvature are now given relative to the spontaneous curvature imposed by the protein density distribution. 
The third contribution from the bending rigidity is no longer weighted with the membrane curvature but with its deviation from the protein-imposed preferred curvature. 
The final term couples the curvature tensor to the gradients in the protein distribution.
Notably, upon defining the deviatoric stress tensor
\begin{equation}
     \Sigma^{D}_{ab} = \frac{\chi}{2} \left[ \frac{1}{2}(\partial_c \phi)(\partial_d \phi)g^{cd}g_{ab} - (\partial_a \phi)(\partial_b \phi) \right] \,,
\end{equation}
one can rewrite the equation as
\begin{align}\label{eq:dlessheq}
    \frac{1}{\sqrt{g}\,\gamma }
    \, \partial_t h 
    = \,
    & 
    \big[ 
    f (\phi) + e(\phi,C)
    \big] \, C 
    - \kappa \nabla^2 (C - C_0 \phi) 
    \notag \\ 
    &+
    \kappa \, (C - C_0 \phi) \, (-C^2 + 2 C_\text{G})
    \notag \\
    &+ 
    C^{ab} \, \Sigma^D_{ab} 
    \, . 
\end{align}
Now the first term contains only the local free energy contributions and the gradients in the protein density are absorbed into the deviatoric stress tensor (last term).
To leading order in the curvature, this equation reduces to
\begin{equation}
    \frac{1}{\sqrt{g} \, \gamma}
    \, \partial_t h 
    = \sigma(\phi) \, C 
    - \kappa \nabla^2 (C {-} C_0 \phi) 
    +
    C^{ab} \, \Sigma^D_{ab} \,,
\end{equation}
with an effective tension term
\begin{align}
    \sigma(\phi)
    =
    \sigma 
    + f (\phi)  
    + \tfrac12 \kappa \, C_0^2 \, \phi^2 \, .
    \label{eq:effectivetension}
\end{align}
This effective tension incorporates the original surface tension $\sigma$, a thermodynamic term related to the protein free energy density $f(\phi)$, and one mechanical term that arises from the protein-induced curvature $C_0$. 
These contributions can be interpreted as a type of \textit{surfactant} effect. 
Just as surfactants alter the properties of liquid interfaces by modifying surface tension, these additional terms adjust the membrane's surface energy. Specifically, the proteins' influence can either stabilize or destabilize the membrane, akin to how surfactants can either decrease or increase surface tension.
Finally, linearizing in the height profile with {$g_{ab}=\delta_{ab}+\mathcal{O}(h^2)$} one obtains
\begin{align}
    \frac{1}{\gamma}
    \, \partial_t h 
    {\ =\ }&\sigma(\phi) \, \nabla_\perp^2 h 
    - \kappa \nabla_\perp^2 (\nabla_\perp^2 h {-} C_0 \phi) \nonumber\\
    &+
    \Sigma^D_{ab} \, (\partial_a \partial_b h) 
    \, ,
\label{eq:shape_mem_prot_linear}    
\end{align}
where we assume summation over repeated indices.
For these three terms, we now have the following physical interpretation. 
The first term represents a generalized surface tension, which includes the bare membrane tension, a surfactant-like term due to the local free energy costs for the protein density, and a mechanical term due to the spontaneous curvature induced by the proteins. 
The second term accounts for the bending cost of the membrane relative to the configuration defined by the protein-induced spontaneous curvature profile $C_0 \phi$. 
The third term describes the coupling between the deviatoric stress of the protein density and the curvature tensor of the membrane.

\section{Ostwald ripening and arrested coarsening}
\label{sec:coarsening}

\begin{figure*}[tb]
\centering
\includegraphics[width=\textwidth]{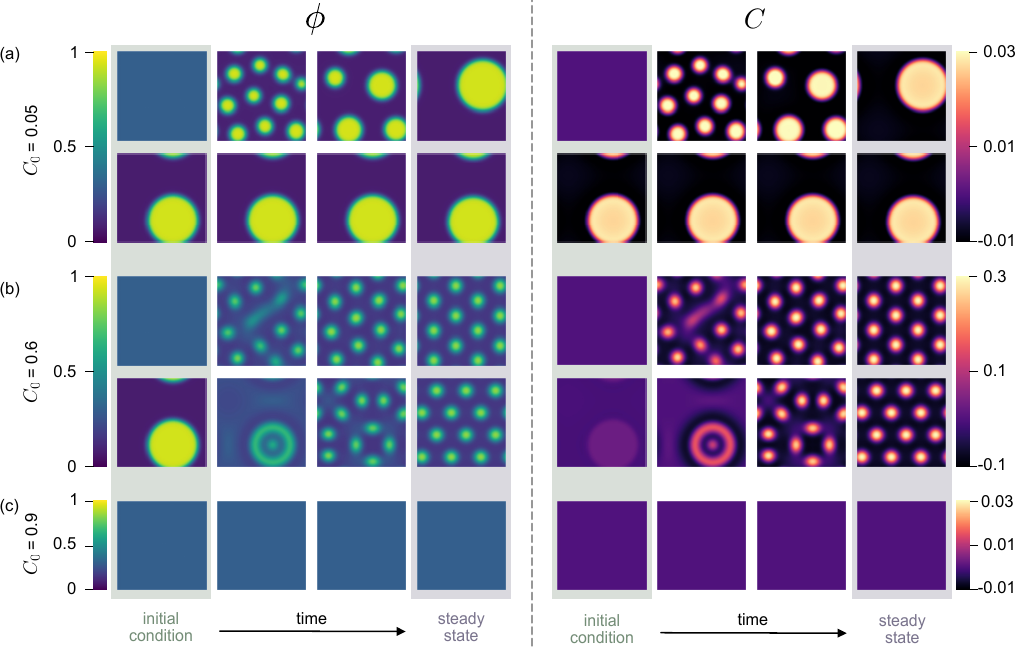} 
\caption{Time evolution of the protein area fraction $\phi(x,y,t)$ (color scale, left) and corresponding membrane curvature $C(x,y,t)$ (color scale, right) for three different protein-induced curvatures ${C_0=0.05}$ (a), ${C_0=0.6}$ (b) and ${C_0=0.9}$ (c). The respective upper and lower time sequences show the dynamics that emerge from two distinct initial conditions: {a protein area fraction} ${\phi_0(x,y) = 0.3+\xi}$ with Gaussian zero-mean white noise $\xi$ of amplitude $\sim\num{5e-4}$ (upper rows), and a single droplet (lower rows). 
The remaining dimensionless parameters are chosen as ${ \sigma = 0.16}$, ${ \gamma = 1}$, ${\kappa = 20}$, and ${\chi = 3}$.      
Spinodal decomposition and coarsening dynamics are observed for ${C_0 = 0.05}$. 
The final steady state is a single droplet.
In contrast, for ${C_0 = 0.6}$, the steady state comprises multiple droplets. 
Starting from a random protein distribution, droplets form by spinodal decomposition, but the coarsening process is eventually arrested.
Moreover, an initial single droplet state is unstable and splits into multiple droplets. Increasing the protein curvature even further leads to a stabilization of the homogeneous state.
}
\label{fig:temp_evo}
\end{figure*} 

In this section we present results from numerical solutions (Appendix~\ref{sec:numericalsim}) of the dynamic equations~\eqref{eq:dyn_phi_nd} and \eqref{eq:dyn_h_nd} to analyze the impact of the mechanical feedback on protein pattern formation and the system's coarsening behavior as a function of the protein-induced curvature $C_0$.
As discussed in the previous section, we choose the remaining parameters such that on a flat, non deformable membrane, the homogeneous steady state ${\phi(x,y)=\bar{\phi}}$ is unstable against small spatial perturbation, i.e., the system is in the spinodal decomposition regime.

For small induced spontaneous curvatures, such as ${C_0=0.05}$, the influence of membrane geometry on the phase separation dynamics of proteins is nearly negligible, resulting in dynamics similar to classical Ostwald ripening described by the Cahn-Hilliard equation; see~Fig.\@~\ref{fig:temp_evo}(a) and supplemental video 1.
After an initial perturbation of the homogeneous steady state ${\phi_0 = 0.3}$, multiple high-density droplets form, and larger droplets subsequently grow at the expense of smaller ones (top row in Fig.\@~\ref{fig:temp_evo}(a)). 
The Ostwald ripening process continues until only a single droplet remains \cite{Lifshitz:1961, Wagner1961Theorie}. 
Consistently, when simulations are initialized with a single droplet, it remains stable over time; see the second row in Fig.~\ref{fig:temp_evo}(a) for ${C_0 = 0.05}$, supplemental video 2.

In contrast, when the protein-induced curvature is increased to ${C_0=0.6}$, the effects of the mechanical feedback become increasingly prominent.
The onset of pattern formation from a homogeneous steady state is similar for ${C_0 = 0.05}$ and ${C_0 = 0.6}$, but is delayed for the larger value of protein-induced curvatures; 
see Fig.~\ref{fig:temp_evo}(b), supplemental video 3. 
Remarkably, in contrast to standard Cahn-Hilliard dynamics, the mechanical feedback arrests the Ostwald ripening process, and a regular pattern with a preferred droplet size emerges as the system's steady state.
To verify that the mechanical deformations induce a length-scale selection, we also performed numerical simulations of the dynamics starting from a single droplet as an initial condition; Fig.\@ \ref{fig:temp_evo}(b), second row and supplemental video 4.
One observes that the initial droplet nearly dissolves into the surrounding low-density phase, giving rise to a pattern with 12 droplets. 
We attribute this to the high cost of membrane deformation associated with large protein aggregates, which leads to a gradual dissolution of a single droplet into several smaller ones.

Further increasing the protein curvature to ${C_0=0.9}$ leads to a stabilization of the homogeneous state; see Fig.~\ref{fig:temp_evo}(c). The initial noise decays and the state with vanishing curvature and a homogeneous protein distribution is stable against small perturbations. Demixing into high and low density phase involves areas with smaller and larger preferred curvature. The energetic cost of smoothly transitioning between these regions increases with $C_0$ and therefore suppresses demixing for $C_0 > C_0^c$.

In summary, the mechanical coupling through protein-induced curvature can suppress demixing and arrest the Ostwald ripening process. Moreover, it facilitates the selection of a well-defined wavelength in the final steady-state patterns. In the following chapters, we use linear stability analysis and thermodynamic arguments to investigate the underlying mechanism.

\section{Linear stability analysis}
\label{sec:LSA}

To better understand the dynamics and the emerging patterns, we perform a linear stability analysis (LSA) of the coupled protein and membrane dynamics, Eq.~\eqref{eq:dyn_phi_nd} and Eq.~\eqref{eq:dyn_h_nd}, around a spatially homogeneous steady state. 
We start with a linear stability analysis of the full dynamic equations and investigate the dispersion relation for varying protein-induced curvature $C_0$. 
In addition, we exploit the thermodynamic nature of the system to obtain analytical expressions for the dispersion relation and the phase boundaries using free energy arguments.

\begin{figure*}
\centering
\includegraphics[width=\textwidth]{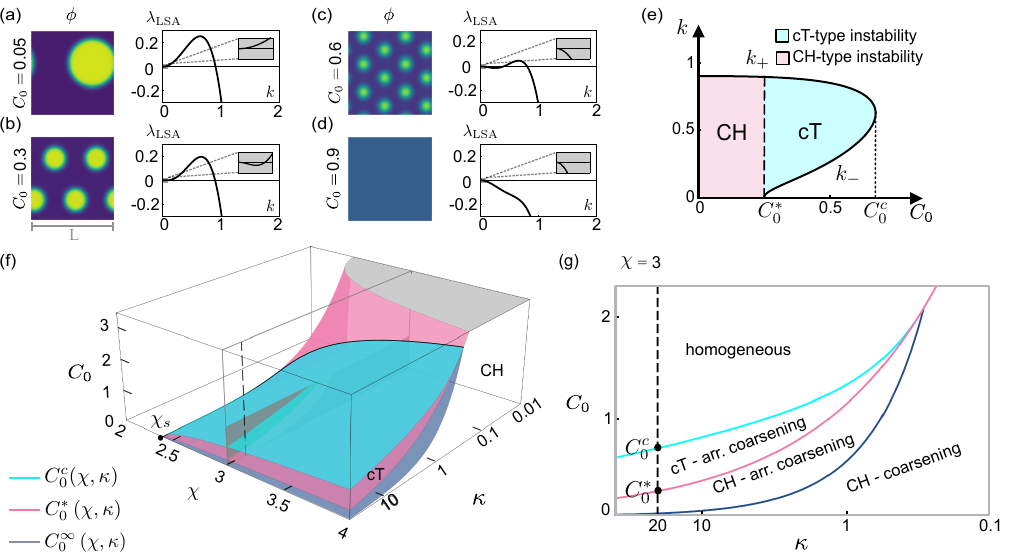} 
\caption{Linear stability analysis;
(a-d) Dispersion relations $\lambda_\mathrm{LSA}(k)$ (right) and the corresponding steady-state density profiles $\phi(x,y)$ (left) for $\chi=3$, $\kappa=20$ and varying protein-induced curvatures ${C_0 = 0.05, 0.3, 0.6,}$ and $0.9$, respectively. Color scale as in Fig.~\ref{fig:temp_evo} for $\phi\in\left[0,1\right]$. Insets show zoom-ins of the dispersion relations at small wave numbers $k \, \in [0,0.1]$. 
(e) Band of unstable modes $(k_-, k_+)$ as a function of the protein-induced curvature $C_0$. 
For ${C_0 < C_0^\text{*}}$, there is a long-wavelength \textit{Cahn-Hilliard-type} (CH-type) instability with the band extending to zero wavevector, ${k_- = 0}$. 
Above the threshold value $ C_0^\text{*}$ (dashed line), the dispersion relation changes from a CH-type long-wavelength instability (magenta) to a \textit{conserved Turing-type} (cT-type) instability, in which the band of unstable modes is bound away from $k=0$, but connects to the marginal mode $k=0$ (cyan).
Above a critical value $ C_0^c$, lateral instabilities are absent (dotted line). 
(f) Bifurcation diagram derived from the dispersion relation of the effective free energy $\lambda$, Eq.~\eqref{eqn:dispersion}, as a function of the protein-induced curvature $C_0$, the FH parameter $\chi$, and the bending stiffness $\kappa$, which shows the boundaries for a CH-type instability (magenta, middle) and a cT-type instability (cyan, upper). Also shown is the boundary between coarsening and arrested coarsening (dark blue, lower) derived from the effective free energy functional ($C_0^\infty$, Eq.~\eqref{eq:transitioncoarsening}). The dashed line indicates the parameters corresponding to (a-d) and the gray plane the parameter ranges of $C_0$ and $\kappa$ corresponding to the simulation results shown in Fig.~\ref{fig:droplet_distance}. (g) Cut plane of the bifurcation diagram (panel (f)) for $\chi =3$.
All other model parameters as specified in Table~\ref{table:parameters}.}
\label{fig:lsa}
\end{figure*} 

\subsection{Linear stability analysis of the dynamic equations}

To perform a linear stability analysis of Eqs.~\eqref{eq:dyn_phi_nd} and \eqref{eq:dyn_h_nd} around a spatially homogeneous state we consider small perturbations $\delta h (\mathbf{x})$ and $\delta \phi(\mathbf{x})$ with respect to a steady state with spatially uniform density and height fields $(\phi_0, h_0)$; the detailed mathematical analysis is given in Appendix~\ref{append:LSA}. 

Figure~\ref{fig:lsa}(a-d) display the corresponding dispersion relations  $\lambda_{\text{LSA}} (k)$, i.e., growth rates as a function of the wavevector $k$, for a set of different protein-induced curvatures $C_0$ and fixed values for ${\chi = 3}$ and ${\kappa = 20}$.
Depending on the magnitude of $C_0$, we observe qualitatively different instabilities.
For small parameter values, e.g., ${C_0 = 0.05}$, the instability is of type-II~\cite{Cross-Hohenberg.1993}, i.e., a long-wavelength instability with a band of unstable modes $(0,k_+)$ extending to zero wavevector.
This is the same type of instability as obtained for the spinodal decomposition regime of the Cahn-Hilliard model in flat geometry (\textit{Cahn-Hilliard-type instability}). 
Above a threshold value ${C_0^*\approx 0.25}$, the dispersion relation becomes qualitatively different and exhibits a band of unstable modes $(k_-,k_+)$ that is bound away from ${k=0<k_-}$ such that long-wavelength modes are stable; note however that the dispersion relation approaches zero in the limit ${k \to 0}$. 
We refer to this type-I~\cite{Cross-Hohenberg.1993} lateral instability as a \textit{conserved Turing-type instability}, following a recent suggestion in Refs.~\cite{Frohoff2023Non,Frohoff2023Nonreciprocal}.
For even larger protein-induced curvature above the threshold value ${C_0^\text{c} \approx 0.68}$, LSA predicts linear stability of the homogeneous steady state as the largest growth rates $\lambda_{\text{LSA}}(k)$ for all modes are non-positive, Fig.~\ref{fig:lsa}(d). 
We attribute this to the stabilizing effect of the mechanical feedback.
Since curvature mismatches are always associated with energetic costs, these costs must exceed the free energy gain from the protein mixing entropy at a certain threshold value of protein-induced curvature, so that the homogeneous steady state becomes energetically favorable. The linear stability analysis already reveals the instability types but an intuitive interpretation of the dispersion relation is challenging and it does not provide insights into the long-term dynamics. In the next section, we will derive an effective free energy that enables both.

\subsection{Effective free energy in the weakly bending limit and linear stability analysis}
\label{subsec:effectivefreeenergy}

Since we are considering the dynamics of a system relaxing into thermodynamic equilibrium, we can take advantage of the fact that the equilibrium steady state is encoded in the free energy functional.
Our starting point is the free energy functional~\eqref{eqn:freeenergyfunc}, together with the corresponding expressions for the local free energies for the protein density $\phi$ and the membrane height undulations $h$, Eq.~\eqref{eq:free_energy_density_FH} and Eq.~\eqref{eq:free_energy_density_CH}, rewritten in their non-dimensional form.

\subsubsection*{Weakly bending approximation}
If one assumes that the gradients in the height fluctuations of the membrane are weak, the surface element can be approximated as
\begin{align}
	\mathrm{d}A 
	\approx 
    \mathrm{d} x \, \mathrm{d} y \,
    \big( 
    1+\tfrac{1}{2}|\nabla_\perp h|^2
    \big)
	\, .
\end{align}
Note that this yields a surface tension term for the membrane of the form $\frac12 \sigma \int \mathrm{d}^2 x \, (\nabla_\perp h)^2$ but in the same way also multiplies all the terms in the Flory-Huggins free energy. This has, as we will see below, important implications for the equilibrium steady state patterns since it gives rise to an effective surface tension.
Moreover, we approximate the membrane curvature by its leading order term
\begin{equation}
    C 
    \approx 
    \nabla_{\perp}^2 h
    \, ,
\end{equation}
where ${\nabla_\perp^2 = \partial_x^2 + \partial_y^2}$. 
Taken together, this weakly bending approximation results in a free energy functional which is quadratic in the height fluctuations
\begin{align}\label{eq:quadraticF}
    \mathcal{F}
    = 
    &\int \! \mathrm{d}^2 x \,
    \Big\{
    \tfrac12 \kappa \big( \nabla_\perp^2 h - C_0 \phi \big)^2 \nonumber \\
    &+\tfrac12 \Big[ \sigma + f(\phi) +  \tfrac12 \kappa C_0^2 \phi^2 +  \tfrac14 \chi (\nabla_\perp \phi)^2  \Big] (\nabla_\perp h)^2   \nonumber \\
    &- \left(\partial_xh\partial_x\phi + \partial_y h \partial_y\phi\right)^2 + f(\phi) + \tfrac14 \chi (\nabla_\perp \phi)^2 \Big\} \nonumber \\
    &= \mathcal{F}_\mathrm{bend} + \int \! \mathrm{d}^2 x \,
    \Big\{f(\phi) + \tfrac14 \chi (\nabla_\perp \phi)^2 \Big\} \, .
\end{align}

To proceed further, we assume that in the terms proportional to $(\partial_ah)(\partial_bh)$ one can replace the protein density field $\phi$ by its spatial average $\bar{\phi}$ such that the bending free energy becomes quadratic in both the protein density and the height field
\begin{align} 
    \mathcal{F}_\text{bend}
    \approx 
    &\int \! \mathrm{d}^2 x \,
    \Big\{
    \tfrac12 \kappa \big( \nabla_\perp^2 h - C_0 \phi \big)^2  \nonumber \\
    &+\tfrac12 \Big[ \sigma + f(\bar{\phi}) +  \tfrac12 \kappa C_0^2 \bar{\phi}^2  \Big] (\nabla_\perp h)^2   
    \Big\}  \, .
\end{align}
Then, the bending free energy functional can be written in Fourier space as follows:
\begin{align}
    \mathcal{F}_\text{bend}
	= 
	\sum_k 
	\Big\{
    &
    \tfrac{1}{2}
    \big[ 
    \kappa k^4 
    + 
     \sigma(\bar{\phi}) k^2
	\big] \, 
	|h_k|^2 
    \nonumber \\
    +&\tfrac{1}{2}
    \kappa C_0^2  
    |\phi_k|^2 
    + 
    \kappa C_0 k^2 
    h_k \phi^*_k
    \Big\} \, ,
\label{eq:bend_harmonic_form}    
\end{align}
with the effective tension
\begin{equation}
\label{eq:effective_tension}
    \sigma(\bar{\phi}) 
    =
    \sigma + f(\bar{\phi}) + \tfrac{1}{2} \kappa \, C_0^2 \bar{\phi}^2 \, .
\end{equation}
The two additional terms, other than the membrane surface tension $\sigma$, arise from the changed free energy costs or gains associated with protein coverage of the membrane (second term) and the induced spontaneous curvature (third term). We focus on parameter regimes where this effective surface tension remains positive, as a negative value would signify an instability in the overall system.

Given the harmonic form of the coupling term in Eq.~\eqref{eq:bend_harmonic_form}, one can perform the path integral in the partition sum over the bending modes by completing the square. 
Thereby, one obtains an effective coupling term of the form 
\begin{align}
    \mathcal{F}_\text{coupling}
  	= 
  	\frac12 
  	\sum_k 
  	\bigg[ \kappa  C_0^2 -
  	\frac{ (\kappa \, C_0)^2 k^2 }
  	{\sigma(\bar{\phi})+ \kappa  k^2}
  	\bigg] 
  	|\phi_k|^2 \, ,
\end{align}
where the summation runs over all Fourier modes compatible with the chosen boundary conditions (typically chosen as periodic boundary conditions) and the system size $L$.
Combining this effective coupling term with the remainder of the FH free energy (see Eq.~\eqref{eq:quadraticF}) gives us the following effective total free energy functional
\begin{align}
    \mathcal{F}_\text{eff} 
    = 
    \frac12 
  	&\sum_k 
  	\bigg[ \,  
    \kappa C_0^2 -
  	\frac{(\kappa  C_0)^2 k^2 }
  	{\sigma(\bar{\phi})+ \kappa k^2} 
    +  \tfrac12 \chi k^2 
    \bigg] \, 
  	|\phi_k|^2 
   \nonumber \\
    + 
    &\int \mathrm{d}^2 x \ f (\phi)  \, .
\label{eq:eff_free_energy}
\end{align}
Importantly, the effective coupling term, which is due to the long-range nature of the membrane-mediated interactions, changes the physics at large length scales.
In contrast to the original stiffness term $\frac14 \chi \, k^2 |\phi_k|^2$ in the FH free energy, it behaves differently at long wavelengths.
Thus, for a finite surface tension $\sigma (\bar{\phi})$, one finds asymptotically in the long wavelength limit, $\lvert k\rvert\ll\sigma(\bar\phi)/\kappa$,
\begin{align}
    \mathcal{F}_\text{coupling}
  	\approx 
  	\frac12 
  	\sum_k 
 	\bigg[ \kappa C_0^2 - \frac{(\kappa \, C_0)^2}{\sigma (\bar{\phi})}  
  	\, k^2 \bigg]
  	|\phi_k|^2 \, .
\end{align}
This reduces the magnitude of the stiffness term, $\sim k^2$ in $\mathcal{F}_\mathrm{eff}$, and can even make it negative, transforming a Cahn-Hilliard-type instability into a conserved Turing-type instability, as we will discuss next.

\subsubsection*{Linear stability analysis of effective free energy}

With the above form of the effective free energy one can write down the dynamics in the limit where membrane dynamics is 
equilibrated on the timescale of the protein density dynamics as
\begin{equation}
    \partial_t \phi  
    = 
    \nabla^2 \frac{\delta {\cal F}_\text{eff}}{\delta \phi}
    \, .
\end{equation}
Linear stability analysis then gives for the growth rates of the Fourier modes, 
\begin{align}
	\lambda (k) = 
	- k^2 
	\Big[ 
 f'' (\bar{\phi}) + \kappa \, C_0^2 
    + \tfrac12 \chi_\text{eff}(k)\, k^2 
    \Big] 
    \, ,
 \label{eqn:dispersion}
\end{align}
where $\chi_{\text{eff}} (k)$ denotes the stiffness parameter renormalized by the bending modes and the generalized surface tension,
\begin{align}
    \chi_\text{eff} (k) 
    = 
    \chi 
    - 
    \frac{2 \left(\kappa \, C_0\right)^2}{\sigma(\bar{\phi}) + \kappa \, k^2}   
  \, .
\end{align}
In addition, the leading order term $-f''(\bar{\phi}) \, k^2$, which determines the position of the spinodal in the CH model through the sign change of the curvature of the free energy density $f''(\bar{\phi})$, is shifted by a constant $\kappa \, C_0^2$, depending on the spontaneous curvature.

Figure~\ref{fig:lsa}(f) shows the bifurcation diagram resulting from the dispersion relation of the effective model, Eq.~\eqref{eqn:dispersion}, as a function of the FH parameter, bending stiffness, and spontaneous curvature. 
It demonstrates that, depending on the values of these parameters, the spatially uniform state can become unstable through either a CH-type or conserved Turing-type instability. These transitions align with the transitions obtained from the full model independently of the timescales (see Appendix~\ref{append:comparisonLSA}).
In the regime where the instability transition is of CH-type, i.e., a long-wavelength instability, the spinodal line is given by the condition ${f'' (\bar{\phi}) + \kappa \, C_0^2 = 0}$, resulting in 
\begin{equation}
   2 \chi
   = 
   \frac{1}{\bar{\phi}\,(1- \bar{\phi})} + \kappa \, C_0^2 
   \, .
\label{eq:modified_spinodal} 
\end{equation} 
Below this spinodal surface, the dynamics is essentially equivalent to those described by the standard Cahn-Hilliard model.
This involves an initial phase of spinodal decomposition followed by a subsequent coarsening process, see supplemental video 1.

However, this CH-type transition with a long-wavelength instability can be pre-empted by a finite wavelength conserved Turing-type instability, where a band $(k_-,k_+)$ of modes becomes unstable.  
This band emerges if the following condition is met (see Appendix~\ref{append:LSA_FE})
\begin{align}
    C_0^{c}\,^2
    &=  \frac{4}{\bar{\phi}^2 \left( \bar{\phi}^2 \chi -16\kappa \right)} 
    \bigg( \bar{\phi}^2 f'' (\bar{\phi}) +  \sigma^{}_\mathrm{FH} \Big(4 -\frac{1}{2 \kappa} \bar{\phi}^2 \chi \Big) 
    \notag \\
    &- 2 \sqrt{\bar{\phi}^2 f''(\bar{\phi}) \left( 4 \kappa f'' (\bar{\phi})/\chi -2 \sigma^{}_\mathrm{FH} \right) + 4 \sigma_\mathrm{FH}^2}
    \bigg) \, ,
\label{eq:typeI_instability}
\end{align}
where one has ${f''(\bar{\phi}) = 1/[\bar{\phi}(1-\bar{\phi})] - 2 \chi}$ and we defined ${\sigma^{}_\mathrm{FH} := \sigma + f(\bar{\phi})}$. 
The resulting bifurcation surface is depicted as the cyan upper surface $C_0^c (\chi, \kappa)$ in  Fig.~\ref{fig:lsa}(f). 
The boundary between the conserved Turing-type and CH-type transitions is given by equating Eq.~\eqref{eq:modified_spinodal} and Eq.~\eqref{eq:typeI_instability}; see the thick black line in Fig.~\ref{fig:lsa}(f).
For parameter ranges with an initial conserved Turing-type instability, Eq.~\eqref{eq:modified_spinodal} delineates the change from this conserved Turing-type to a CH-type dispersion relation.

Comparing the dispersion relation of the full model, $\lambda_{\text{LSA}}(k)$ with that of the effective model, Eq.~\eqref{eqn:dispersion}, two observations can be made. 
First, the effective dispersion relation is obtained from the full dispersion relation in the limiting case of a strong timescale separation between the protein density dynamics and the membrane dynamics. 
However, in our numerical simulations we chose comparable timescales, i.e.\@ $\gamma=1$, for both dynamics. 
As membrane deformations are rather weak, we still observe a co-localization of deformations with protein aggregates, and hysteresis from previous membrane configurations is still negligible.
Second, the phase boundaries for the Cahn-Hilliard and conserved Turing-type instabilities remain unchanged even if there is no separation of timescales; see Appendix~\ref{append:comparisonLSA}.

\section{Thermodynamics and length-scale selection}
\label{sec:lengt_scale_selection}

To better characterize the emergence of the finite pattern length scale and the arrest of coarsening, we simulated the system dynamics across a larger region in parameter space, for a fixed value of ${\chi = 3}$ and encompassing variations in protein-induced curvature (${C_0 \in [0,1.2]}$) and membrane bending stiffness (${\kappa \in[0,40]}$).
Figure~\ref{fig:4} shows the corresponding results for the inter-droplet distance (see Appendix~\ref{append:distmeasure}) in the stationary state.
It reveals the existence of two distinct phase boundaries. 

First,  a \textit{spinodal line} (cyan) marks the onset of the instability of a spatially uniform state against small perturbations. In the parameter range shown, this transition is a conserved Turing-type instability, given by Eq.~\eqref{eq:typeI_instability}; there is excellent agreement between the predicted and the numerically observed transition (Fig.~\ref{fig:4}).
Below the spinodal line, our numerical simulations show that the system initially undergoes spinodal decomposition and Ostwald ripening.
However, this coarsening process does not reach completion; instead, it eventually arrests, and results in the formation of a spatial pattern with periodic arrangements of droplets of finite size.
This is distinct from the Ostwald ripening process that we observe in a parameter regime where the transition is driven by a Cahn-Hilliard-type long-wavelength instability.
Upon decreasing the protein-induced curvature $C_0$ for a fixed value of the bending rigidity $\kappa$, our simulations show that both the distance between the droplets and their size increase; see the black dashed line and the corresponding panels for the patterns in Fig.~\ref{fig:4}.  
Eventually, in the finite-sized system simulated here, there is a threshold line for the protein-induced curvature below which the system exhibits only a single droplet; see the solid purple line in  Fig.~\ref{fig:4}.

\begin{figure}[tb]
  \centering
    \includegraphics[width=0.5\textwidth]{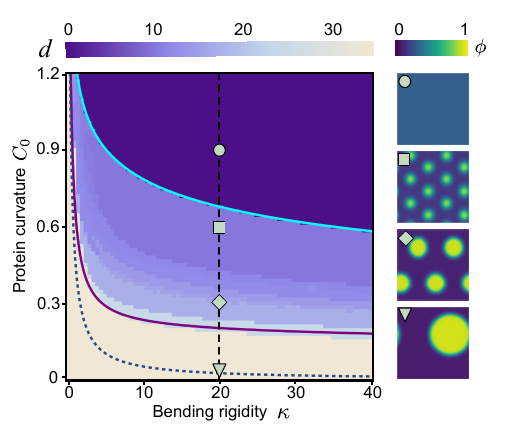}
\caption{Inter-droplet distance $d$ (color code) in the final stationary state of numerical simulations of the system described by Eq.~\eqref{eq:dyn_phi_nd} and Eq.~\eqref{eq:dyn_h_nd}, for different values of protein-induced curvature $C_0$ and bending rigidity $\kappa$ in a square domain with side length ${L = 2\, \si{\micro \meter}}$. 
The spinodal line for the conserved Turing-type transition ($C_0^c$, cyan, top) marks the onset of pattern formation predicted by linear stability analysis; see Eq.~\eqref{eq:typeI_instability}. 
With decreasing protein-induced curvature ($C_0$), the length scale of the emerging droplet pattern increases (black dashed line at $\kappa=20$); the panels on the right show corresponding snapshots of these patterns at the parameters indicated by the green symbols.
The transition from a finite number of droplets to a single droplet in a finite size system is indicated by a solid purple (middle) line ($C_0^L$, Eq.~\eqref{eqn:purpleline}). 
The phase boundary between an infinitely extended system that shows arrested coarsening and one that exhibits Ostwald ripening is shown by the dashed blue (bottom) line ($C_0^{\infty}$, Eq.~\eqref{eq:transitioncoarsening}).
All system parameters as in Fig.~\ref{fig:lsa}, where parameter ranges of $C_0$ and $\kappa$ are indicated as a gray plane.
}
\label{fig:4}
\end{figure}

\subsubsection*{Thermodynamic length-scale selection}

Why does the system exhibit coarsening arrest within an intermediate range of protein-induced curvature and bending rigidity? 
One way to address this question would be to use methods from nonlinear dynamics and perform a weakly nonlinear analysis along the lines introduced by Matthews and Cox~\cite{Matthews:2000}, which accounts for the presence of the long-wavelength marginal mode at ${k=0}$.
In the present case,  however, there is a more straightforward approach, as the dynamics relax to a thermal equilibrium state determined---at the mean-field level---by the minimum of the free energy functional.
In general, thermodynamic systems with only short-ranged interactions show Ostwald ripening with the equilibrium state given by complete phase separation. 
Here, however, we have long-ranged interactions between the proteins mediated by the elastic deformation of the membrane surface. 
It is known that the balance between the short-ranged forces driving phase separation and the long-ranged forces that impose ordering constraints can lead to the formation of patterns.
A classical example are block copolymer melts, where the long-range interaction is mediated by connections between different chemical sequences in the copolymer chain~\cite{Liu:1989}.  
In the present context, there is an intriguing twist to this narrative due to the significance of geometric effects inherent in the metric of the membrane surface.

The length scale of the final pattern can be determined from the effective free energy functional derived in the previous section, Eq.~\eqref{eq:eff_free_energy}. 
Following the approach in Ref.~\cite{Liu:1989}, this length scale is obtained by finding the wave vector that minimizes the effective free energy functional density,
\begin{align}
    k_\text{min}^2 
    = C_0
    \sqrt{\frac{2 \sigma(\bar{\phi})}{\chi}}
    - 
    \frac{\sigma(\bar{\phi})}{\kappa} 
    \, .
\label{eqn:kmin}
\end{align}
This relation shows that thermodynamically, one should have a transition from microphase separation (arrested coarsening) to full coarsening for ${k_\text{min}=0}$, equivalent to
\begin{equation}
    C_0^\infty = \frac{1}{\kappa} \, \sqrt{\frac{\chi \, \sigma(\bar{\phi})}{2}} \, .
    \label{eq:transitioncoarsening}
\end{equation}
Substitution of the effective surface tension, Eq.~\eqref{eq:effective_tension}, and solving for $C_0^{\infty}$ yields
\begin{equation}
    C_0^\infty = \sqrt{\frac{\chi (\sigma + f(\bar{\phi}))}{2\kappa^2 \big( 1 - \chi \bar{\phi}^2 /(4\kappa)  \big)}} 
    \, ;
\end{equation}
see the dashed blue line in Fig.~\ref{fig:4} and the dark blue surface in Fig.~\ref{fig:lsa} \footnote{Note, that the transition between coarsening and arrested coarsening is distinct from the transition between CH-Type and cT-Type instabilities.}.

\subsubsection*{Finite-size effects}

On comparing our numerical simulations with the above thermodynamic results we, have to account for finite size effects. The wavevector $k_\mathrm{min}$ defines the inter-droplet distance
\begin{align}\label{eq:distance}
    d 
    = 
    \frac{2 \pi}{k_\text{min}} 
    \, 
\end{align}
of the steady-state pattern. 

\begin{figure}[!t]
\centering
\includegraphics[width=0.95\columnwidth]{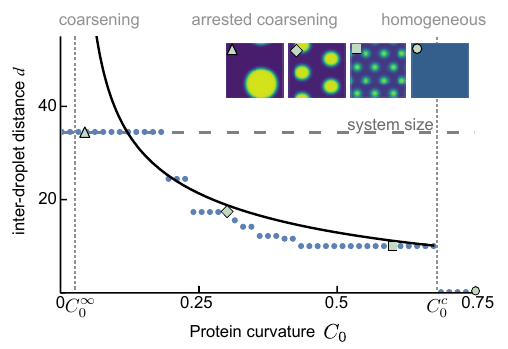}
\caption{ Comparison of the analytical result for inter-droplet distance Eq.~\eqref{eq:distance} (black line) and the simulation results for the cutline ${\kappa=20}$ in Fig.~\ref{fig:4} (blue dots). Both are in good agreement, with minor deviations due to finite size effects and the harmonic approximation used to derive the effective free energy. The panels on top show the pattern corresponding to the length scale indicated by the green symbols. The system size, given by the horizontal dashed line, restricts the maximum possible distance. The regime showing microphase separation is limited by $C_0^\infty$ (Eq.~\ref{eq:transitioncoarsening}) and $C_0^\mathrm{c}$ (Eq.~\ref{eq:typeI_instability}) as indicated by the vertical dashed lines.
}
\label{fig:droplet_distance}
\end{figure} 

Figure~\ref{fig:droplet_distance} compares the simulation data for the cutline ${\kappa = 20}$ in Fig.~\ref{fig:4} alongside this analytical expression for the inter-droplet distance. The average distance between droplets decreases with increasing protein curvature $C_0$, and we observe good agreement with the simulation results. However, as we consider a finite-size system, the inter-droplet distance takes on discrete values. 
Moreover, since the hexagonal symmetry of the periodic arrangement of the droplets is not consistent with the square domain used in the simulations, there are geometric frustration effects; for a discussion, see Appendix~\ref{append:distmeasure}.
Finally, there may also be deviations between the simulation results and the thermodynamic expression because we have used harmonic approximations in the derivation of the effective free energy (Sec.~\ref{subsec:effectivefreeenergy}).

Figure~\ref{fig:droplet_distance} also illustrates that the inter-droplet distance is limited by the system size, and the boundary between coarsening and arrested coarsening is shifted compared to an infinite system. The maximum distance between two droplets in our finite domain is ${d_2=L/\sqrt{2}}$. Thus, we estimate the boundary between microphase separation (two or more droplets) and complete Ostwald ripening (one droplet) for our finite square domain with length $L$ through the condition
\begin{align}
    d = \frac{L}{\sqrt{2}} \, . 
\label{eqn:purpleline}
\end{align}
The corresponding critical value of the protein-induced curvature $C_0^{L}(\kappa)$ is shown in Fig.~\ref{fig:4} as the purple line. As the system size increases and finite-size effects decrease, we expect the agreement between the analytically predicted and numerically measured pattern length scales to improve even further, in addition to the already good agreement for the finite-size system. Specifically, as the systems become larger, we anticipate that the critical transition lines will converge, i.e., $C_0^L \rightarrow C_0^{\infty}$ in the limit $L\rightarrow\infty$. Since we do not anticipate any fundamentally new effects to arise at larger scales, we will omit the technical challenge of performing numerical simulations of larger systems here.

\section{Discussion}
\label{sec:discussion}

In this study, we investigated phase separation on membranes that dynamically evolve their shape by integrating a Flory-Huggins theory for symmetric binary mixtures with a Canham-Helfrich theory for fluid elastic membranes.
Specifically, we focused on systems where the density of molecules undergoing phase separation induces spontaneous membrane curvature, thereby facilitating a coupling with the membrane's mechanical deformations.

We employed a general covariant framework to describe phase-separation dynamics on membranes with dynamically evolving geometry, deliberately avoiding the use of small deformation or dilute phase expansions that are commonly used to simplify the analysis. 
The resulting set of coupled dynamic equations for protein density and membrane conformations account for the effects of spatiotemporal variations in the prevailing surface metric, ensure mass conservation, and capture the mechanical coupling arising from protein-induced curvatures.

The analysis of these dynamic equations shows that liquid phase separation on deformable membranes exhibits three qualitatively different phenomenologies.
First, we observe a regime with stable, spatially homogeneous steady states, where proteins maintain a mixed state with a uniform density across a flat membrane.
Second, we find a regime characterized by a fully coarsened phase-separated steady state, where proteins aggregate into a single high-density droplet surrounded by a low-density phase.  
Finally, we also find a regime of arrested coarsening, where protein aggregation is counteracted by the energetic cost of membrane deformation induced by the mechanical coupling to the protein density. 
In this regime, the length scale of the emergent pattern is determined by the trade-off between the thermodynamics of protein mixing and membrane bending energy costs.

An interesting and notable feature of phase separation on deformable membranes is that the dispersion relation changes from a standard Cahn-Hilliard-type long-wavelength instability to a conserved Turing-type instability above a certain threshold value for the spontaneous curvature.
If the system becomes unstable through a Cahn-Hilliard instability, we observe Ostwald ripening, where smaller droplets dissolve into larger ones, resulting in coarsening of the phase-separated structures over time. 
In contrast, the conserved Turing instability, characterized by a band of unstable modes and a marginal mode at zero wavevector in the dispersion relation, drives the formation of spatial patterns with a finite wavelength. 
Since we study the dynamics of a protein-membrane system that relaxes to a thermodynamic equilibrium, one can take advantage of the fact that the thermal equilibrium state is encoded in the free energy functional. 
By minimizing this effective free energy density, where membrane conformations were integrated out, we were able to determine the length scale of the patterns as well as the boundary between coarsening and arrested coarsening. 
The analytical results obtained through this method showed excellent agreement with our numerical simulations.

Similar to recent findings in Ref.~\cite{yu2023pattern,Mahapatra:2021}, our results highlight that curvature-mediated interactions can strongly influence the phase separation dynamics. Ref.~\cite{Mahapatra:2021} primarily focuses on the influence of the geometry on the initial phase separation dynamics and assumes mechanical equilibrium for the membrane. In contrast to that, we present a covariant framework that also incorporates effects arising from a dynamically evolving geometry, which ensures mass conservation as well as advection of the density along the material points during the deformation. Additionally, we highlight that these long-range membrane-mediated interactions, in conjunction with phase separation, can serve as a minimal motif for pattern formation with a well-defined length scale, determined by the material parameters of the system.
Notably, this mechanism does not depend on complex biochemical pathways but rather on generic features of the membrane's lipid composition~\cite{yu2023pattern} or protein interaction with the membrane~\cite{Lipowsky:2022,Christ2020Active}.
The dynamics of the membrane and the density field discussed in our work are qualitatively similar to those in Ref.~\cite{yu2023pattern}, but differ in several key aspects. 
First, we avoid expanding the membrane dynamics beyond the intrinsic limits of Monge representation. We introduce the material derivative to ensure that protein density advection correctly follows membrane deformations, securing the full covariance of the theory.
Second, we incorporate the effects of the surface metric in the Flory-Huggins free energy, resulting in an effective membrane tension. 
Finally, we account for the metric in the time derivative of protein density ensuring accurate protein mass conservation.
These differences lead to distinct predictions for the phase diagram, particularly in how the phase boundaries depend on model parameters such as the protein-induced curvature. 

In our current analysis, we have neglected the in-plane lipid flow within the membrane. 
Extending recent work~\cite{Mahapatra:2021}, which incorporates this flow, to the nonlinear Monge regime explored here would be a valuable next step. 
Additionally, another promising research avenue would involve accounting for the fluid flow of the surrounding liquid beyond the Rouse approximation used in our present study. Moreover, here, we have focused on the effects of an isotropic induced spontaneous curvature. For future studies one could extend the covariant framework to explore the impact of an anisotropic induced curvature or alternative mechanical couplings, such as protein dependent bending rigidity or surface tension. Such general systems exhibiting mechanical interactions can give rise to overhang formation and budding~\cite{Hassinger:2017}. Here we restrict our analysis to a regime where a Monge parametrization is valid. However other parametrizations of Eq.~\eqref{eqn:h_eq} and Eq.~\eqref{eqn:phi_eq} can be used to study these.

Finally, an intriguing open research question that could be explored by adapting our theoretical framework is how dynamics and steady states are affected in systems with broken detailed balance.
This includes systems where the de-mixing dynamics is described by the nonreciprocal Cahn-Hilliard equations~\cite{Schuler2014Spatio,Saha2020Scalar,Frohoff2023Non,Brauns2024Nonreciprocal} and various pattern-forming systems~\cite{Radszuweit2013Intracellular,Edelstein-Keshet2013From,Halatek:2018:Rethinking,Brauns2020Phase,Wuerthner:2023}. 
For instance, two-component mass-conserving reaction-diffusion system on deformable surfaces, exhibit geometrically induced pattern-forming instabilities and the occurrence of oscillations and traveling waves, not present without geometric coupling~\cite{Wuerthner:2023}. 
An extension of the presented work towards weakly driven out-of-equilibrium pattern forming systems may yield additional insights into the interplay between the equilibrium processes of demixing and mechanically mediated length-scale selection and actively driven chemical reactions.
The general correspondence between Cahn-Hilliard models and reaction-diffusion systems with conservation laws~\cite{Bergmann2019System,Brauns:2021} promises the possibility of directly applying the approaches presented here to such systems and extending them to even more general active matter.

\acknowledgments{
We thank Simon Bauer, Tom Burkart, Tobias Roth, and Jan Willeke for stimulating discussions and Padmini Rangamani, Andrej Ko\v{s}mrlj, and Qiwei Yu for their insightful comments on the manuscript.
We acknowledge financial support by the German Research Foundation (DFG) through the Excellence Cluster ORIGINS under Germany's Excellence Strategy (EXC 2094 -- 390783311), by the European Union (ERC, CellGeom, project number 101097810), and the Chan--Zuckerberg Initiative (CZI).
}

\appendix
\section{Non-dimensionalized dynamic equations}
\label{App:Derivation}

As discussed in the main text, we base our study on the non-dimensionalized effective free energy functional
\begin{align}
\label{eqn:nondim_energy}
    \mathcal{F}  = \int \mathrm{d} A 
    \Big[  &\sigma + 
    \frac{\kappa}{2} \big(  
    C - C_0 \phi
    \big) ^2
    \nonumber \\ 
    &+  
    \big(
    \phi \ln{\phi} + (1-\phi) \ln{(1-\phi)}
    \big) 
    \notag  \\ & 
    +
    \chi  \phi (1-\phi) 
    + 
    \frac{\chi}{4} 
    \vert \nabla \phi \vert ^2
    \Big]
    \, .
\end{align}
We assume in the following that all quantities are non-dimensionalized as discussed in Sec.\ref{subsec:nondimensionalization}.
In order to derive dynamic equations for the membrane height field $h$ and the protein density $\phi$, we assume gradient dynamics towards minima of the free energy functional $\mathcal{F}$. 
In particular, we assume relaxational (model A) dynamics for the membrane conformation ${\mathbf{r} = (x,y,h(x,y))^T}$,
\begin{align}
    \mathcal{D}_t \mathbf{r} 
    &= 
    - \gamma \,
    \frac{\delta \mathcal{F}}{\delta \mathbf{r}} \,,
\end{align}
and Cahn-Hilliard (model B) dynamics for the conserved protein field $\phi$,
\begin{align}
    \frac{1}{\sqrt{g}} \, 
    \partial_t\left(\sqrt{g}\phi\right)
    =
    \nabla_a \Big[ \nabla^a  \frac{\delta\mathcal{F}}{\delta\phi} + v_\mathrm{coord}^a\phi \Big]\,.
\end{align}

To derive the explicit form of the dynamic equations, we need to perform the variation of the free energy functional with respect to the membrane parametrization $\mathbf{r}$ and the protein area fraction $\phi$. 
For the dynamics of the membrane conformation we can restrict to normal variations, since we focus on deformations and neglect the in-plane lipid flow within the membrane
\begin{align}\label{eq:timedeppositionvector}
    \mathcal{D}_t \mathbf{r} 
    &= 
    - \gamma \,
    \frac{\delta \mathcal{F}}{\delta r_n}\mathbf{n} \nonumber \\
    &= v_n \mathbf{n} \,,
\end{align}
where $\mathbf{n} = \tfrac{1}{\sqrt{g}}(-\partial_x h,-\partial_y h,1)^T $ is the normal vector of the membrane surface and $v_n$ the normal velocity. We make use of the fact that the variations of the curvature, metric, and metric tensor, respectively, are given by~\cite{Deserno:2014}
\begin{align}
    &\delta C = \left(\Delta + C^2 - 2 C_\mathrm{G}\right) \,\delta r_n \, ,\label{eqn:normvarc}\\
    &\delta g = - 2 \, g \, C \, \delta r_n \, ,\label{eqn:normvarg}\\
    &\delta g^{ab} = 2 \, C^{ab} \, \delta r_n \label{eqn:normvargab}
    \, .
\end{align}
Here 
\begin{equation}
    {C}_{ab} = \frac{1}{\sqrt{1+(\nabla_{\perp} h)^2}} \begin{pmatrix}
    \partial_{x}^2 h & \partial^2_{yx}h \\
    \partial^2_{xy} h & \partial^2_{yy} h
    \end{pmatrix}
\end{equation}
denotes the extrinsic curvature tensor, $C$
is the total curvature, defined in the main text, Eq.\eqref{eq:curvature_monge}, and
\begin{equation}
    {C}_\text{G} = \frac{\det[\nabla_\perp \nabla_\perp h]}{(1+ (\nabla_{\perp} h)^2)^2}
\end{equation}
is the Gaussian curvature. 

Given the free energy functional $\mathcal{F}$, Eq.~\eqref{eqn:nondim_energy}, we find for the local part of the functional 
\begin{align}
    \delta \int \mathrm{d}A \, f_\text{tot} [\phi, g_{ab}, C]
    &= \delta \int \mathrm{d}x \mathrm{d}y \sqrt{g} \, f_\text{tot} [\phi, g_{ab}, C]
    \notag \\
    &= \int \mathrm{d}x \mathrm{d}y \left[ \frac{1}{2 \sqrt{g}} \delta g \, f_\text{tot} + \sqrt{g} \, \delta f_\text{tot}\right] \notag \\
    &= \int \mathrm{d}A \left[ -f_\text{tot} C \, \delta r_n + \delta f_\text{tot} \right]
    . 
\end{align}
Here the total free energy density is given by the sum of the Canham-Helfrich and the Flory-Huggins free energy Eq.~\eqref{eqn:nondim_energy}: 
\begin{align}
    f_{\text{tot}} [\phi, g_{ab},C]
    &= f [\phi, g_{ab}] + e[\phi, C] 
    \\
    &= f\left(\phi\right) + \frac{\chi}{4}
    (\partial_a\phi) 
    (\partial_b \phi) 
    g^{ab} 
    + e (\phi, C) \notag \, ,
\end{align}
where we have used ${\left| \nabla \phi \right|^2 = (\partial_a\phi)( \partial_b \phi) g^{ab}}$. 
By combining the variation of the curvature $C$ (Eq.~\eqref{eqn:normvarc}) and the metric tensor $g_{ab}$ (Eq.~\eqref{eqn:normvargab}) with the above results, one obtains  the variation of the free energy functional 
\begin{widetext}
\begin{align}
    \delta \mathcal{F} 
    &= 
    \int \mathrm{d}A 
    \Big[ 
    - C 
    \big(
    f\left[\phi, g_{ab}\right] + e\left[\phi, C\right] 
    \big) 
    \delta r_n 
    +  
    \frac{ \chi}{4}
    (\partial_a\phi) (\partial_b \phi)
    \delta  g^{ab} 
    + 
    \partial_C e\left[\phi, C\right] \delta C \Big] 
    \notag \\
    &= 
    \int \mathrm{d}A 
    \Big[ 
    - C 
    \big(
    f\left[\phi, g_{ab}\right] + e\left[\phi, C\right] 
    \big) 
    \,\delta r_n 
    +  
    \frac{ \chi}{2}
    (\partial_a\phi) 
    (\partial_b \phi)
    C^{ab} \, \delta r_n  
    + 
    \partial_C e\left[\phi, C\right] 
    \big(
    \Delta + C^2 - 2 C_\mathrm{G}
    \big) 
    \, \delta r_n 
    \Big] 
    \notag \\
    &= 
    \int \mathrm{d}A 
    \Big[ 
    - C 
    \big(
    f\left[\phi, g_{ab}\right] + e\left[\phi, C\right]
    \big) 
    +  
    \frac{ \chi}{2}
    (\partial_a\phi)
    (\partial_b \phi)
    C^{ab}  
    + \Delta
    \partial_C e [\phi,C] 
    + 
    (\partial_C e\left[\phi, C\right]) 
    \big( 
    C^2 - 2 C_\mathrm{G}
    \big) 
    \Big] \, \delta r_n
    \, .
\end{align}
In the last step we have integrated the term $\left(C-C_0 \phi\right) \Delta \delta r_n $ twice by parts and assumed vanishing boundary terms.
Insertion of the equations for the free energy densities gives the normal variation of the free energy functional
\begin{align}
    \frac{\delta \mathcal{F}}{\delta r_n} 
    = 
    &- C \,
    \bigg[ 
    \phi \ln \phi 
    + (1-\phi) \ln (1-\phi) 
    + \chi \phi (1-\phi)
    +
    \frac{\chi}{4}
    \vert \nabla \phi \vert ^2
    + \frac{\kappa}{2}
    \left( C - C_0 \phi \right)^2 
    + \sigma 
    \bigg] 
    \notag \\
    &+ \frac{\chi}{2} 
    \left(\partial_a \phi\right) \left( \partial_b \phi\right) \, C^{ab} + \Delta \kappa (C - C_0 \phi)
    +
    \kappa (C - C_0\phi) 
    \left(C^2
        - 2 C_\text{G}
    \right).
\end{align}
Together with Eq.~\eqref{eq:timedeppositionvector} this defines the time evolution of the height field 
\begin{align}
   \partial_t h &= \sqrt{g} \,  v_n \nonumber \\
    &=  \gamma \, \sqrt{g} \Bigg[
    \big[ 
    f [\phi, C] + e(\phi,C)
    \big] \, C 
    - \kappa \nabla^2 (C - C_0 \phi) 
    +
    \kappa \, (C - C_0 \phi) \, (-C^2 + 2 C_\text{G})
    - 
    \frac{\chi}{2} \, 
    C^{ab} \, (\partial_a \phi) (\partial_b \phi) \Bigg] \, 
\end{align} 
and the coordinate velocities 
\begin{align}
\begin{split}
    v_\mathrm{coord}^1 =  \frac{v_n \, \partial_x h}{\sqrt{g}} \, ,
   \end{split}
   \begin{split}
    v_\mathrm{coord}^2 =  \frac{v_n \, \partial_y h}{\sqrt{g}} \, .
   \end{split}
\end{align}
Analogously, the dynamic equation for the protein density reads
\begin{align}
    \frac{1}{\sqrt{g}} \partial_t (\sqrt{g} \phi) 
    =& \nabla_a \bigg[ \nabla^a\left( \partial_{\phi} e\left(\phi, C\right)+\partial_{\phi}f\left(\phi\right) -\frac{\chi}{2}\nabla^2\phi \right) + v_\mathrm{coord}^a \phi \bigg] \nonumber \\
    =&
    \nabla_a 
    \bigg[ \nabla^a \left(
        -{C}_0 {\kappa}
        ({C} - {C}_0 \phi)
        +
        \ln \bigg(\frac{\phi}{1-\phi}\bigg)
        +
        {\chi}
        (1 - 2 \phi)
        -
         \frac{{\chi}}{2}
         {\nabla}^2 \phi \right) + v_\mathrm{coord}^a \phi
    \bigg]\,.
\end{align}

\end{widetext}

\section{Numerical Simulations}
\label{sec:numericalsim}

We solved Eq.~\eqref{eq:dyn_phi_nd} and Eq.~\eqref{eq:dyn_h_nd} numerically in two spatial dimensions using finite element methods with the commercially available software COMSOL Multiphysics v.6.1~\cite{COMSOL}. 
The simulations were performed on a square  domain with side length $L = 2\, \si{\micro \meter}$ and periodic boundary conditions. 
As an initial condition, we used one droplet, as specified in the provided COMSOL file \cite{supplemental_notebook}, or a homogeneous protein density $\phi_0(x,y)=0.3$ perturbed by Gaussian zero-mean white noise with an amplitude $\sim\num{5e-4}$. The total protein mass was chosen to be the same for the one droplet and the homogeneous protein density.

\section{Linear stability analysis}
\label{append:LSA}

We perform a linear stability analysis (LSA) around the spatially homogeneous state ($h_0,\bar{\phi}$). 
To this end, we introduce small perturbations of the density field ${\phi = \bar{\phi} + \delta \phi}$ and the height field ${h = h_0 + \delta  h}$ with respect to the spatially homogeneous state and linearize the dynamic equations, Eq.~\eqref{eq:dyn_phi_nd} and \eqref{eq:dyn_h_nd},
\begin{align}
    \partial_t 
    \begin{pmatrix}
        \delta h \\
        \delta \phi
    \end{pmatrix}
    = J \cdot 
     \begin{pmatrix}
        \delta h \\
        \delta \phi
    \end{pmatrix}
    \end{align}
   with the Jacobian
\begin{align}
J =
\begin{pmatrix}\label{app_eq:j}
    J_{hh} & J_{h\phi} \\
 J_{\phi h} & J_{\phi\phi} \\
\end{pmatrix},
\end{align}
where
\begin{subequations}\begin{align}
    J_{hh} 
    &= -\frac{1}{2} {\gamma}  {k}^2 \big[2 \bar{\phi} \left({\chi}-\ln \left(1-\bar{\phi}\right)+\ln \bar{\phi} \right) \\
    &\quad \,+2 \left({\kappa}  {k}^2+{\sigma} +\ln \left(1-\bar{\phi}\right)\right) + \bar{\phi}^2 \left({\kappa}  {C}_0^2-2 {\chi}\right)\big] \, , \nonumber \\
    J_{h\phi} &= {\gamma}  {\kappa}  (-{C}_0) {k}^2  \, , 
    \\
    J_{\phi\phi} &= -\frac{{k}^2}{2 \left(\bar{\phi}-1\right) \bar{\phi}} \big[\bar{\phi}^2 \left({\chi} \left(k^2-4\right)+2 {\kappa}  {C}_0^2\right) 
    \\
    &\quad \, -\bar{\phi} \left({\chi} \left({k}^2-4\right)+2 {\kappa}  {C}_0^2\right)-2\big] \, 
    \nonumber \\
    J_{\phi h} &= {\kappa}  (-{C}_0) {k}^4
    \, .
\end{align}
\end{subequations}
The largest eigenvalue of the Jacobian determines the highest growth rate.  
The analysis of analytical expressions is implemented in Mathematica 13.1, and the corresponding notebook is available at~\cite{supplemental_notebook}.
Solving for the largest eigenvalues of the Jacobian $J$ as a function of wavevector $k$ yields the dispersion relations and the bifurcation diagram shown in Fig.~\ref{fig:lsa}. 

\section{Linear stability analysis for the reduced model with an effective free energy}
\label{append:LSA_FE}
In this chapter we analyze the dispersion relation obtained from the effective free energy functional, Eq.~\eqref{eqn:dispersion}.
We determine the region in the $(\chi,  \kappa,  C_0)$ phase space, where a conserved Turing-type instability exists, by investigating the roots $ k _- \ \text{and} \  k_+$ of the dispersion relation Eq.~\eqref{eqn:dispersion}
\begin{widetext}
\begin{subequations}
\begin{align}
    k_+ 
    &= \frac{1}{\sqrt{\chi \kappa}}\bigg[ - ( \tfrac12 \chi \sigma\left(\bar\phi\right) + \kappa f''({\bar\phi}) )  
    + \sqrt{(\tfrac12 \chi \sigma\left(\phi\right) + \kappa f''({\bar\phi}))^2 - 2 \chi\kappa\sigma\left(\bar\phi\right)(f''(\bar\phi) + \kappa C_0^2)}\bigg]^{\frac{1}{2}} , \label{eq:kplus} \\
    k_- 
    &= \frac{1}{\sqrt{\chi \kappa}}\bigg[ - ( \tfrac12 \chi \sigma\left(\bar\phi\right) + \kappa f''({\bar\phi}) ) 
    - \sqrt{(\tfrac12 \chi \sigma\left(\bar\phi\right) + \kappa f''({\bar\phi}))^2 - 2 \chi\kappa\sigma\left(\bar\phi\right)(f''({\bar\phi}) + \kappa C_0^2)}\bigg]^{\frac{1}{2}} , \label{eq:kmnius}  
\end{align}
\end{subequations}
A conserved Turing-type instability can only exist in the regions where $ k_-$ is defined. 
From this, two conditions can be derived
\begin{subequations}
\begin{align}
    0 
    &= - ( \tfrac12 \chi \sigma\left(\bar\phi\right) + \kappa f''({\bar\phi}) ) 
    - \sqrt{(\tfrac12 \chi \sigma\left(\bar\phi\right) + \kappa f''({\bar\phi}))^2 - 2 \chi\kappa\sigma\left(\bar\phi\right)(f''({\bar\phi}) + \kappa C_0^2)} \, ,
    \label{eqn:condition1} 
    \\ 
    0 
    &= \big(\tfrac12 \chi \sigma(\bar\phi) + \kappa f''({\bar\phi}) \big)^2 
    - 2 \chi\kappa\sigma (\bar\phi) 
    \big(f''({\bar\phi}) + \kappa C_0^2 \big)
    \, .
    \label{eqn:condition2}
\end{align}
\end{subequations}
We recover the critical protein induced-curvature $C_0^{*}(\chi , \kappa)$ from the first equality, Eq.~\eqref{eqn:condition1}. However, this solution is only valid if the condition ${ \tfrac12 \chi \sigma\left(\bar\phi\right) + \kappa f''({\bar\phi}) \leq 0}$ is fulfilled. 
The black solid line in Fig.~\ref{fig:lsa}(f) depicts the boundary, given by ${\tfrac12 \chi \sigma\left(\bar\phi\right) + \kappa f''({\bar\phi})  = 0}$. 
The second equality is equivalent to the condition ${k _- = k_+}$, which determines the spinodal for the conserved Turing regime
\begin{align}
    C_0^\mathrm{c} &= \bigg[\frac{4}{{\bar\phi}^2 \left( {\bar\phi}^2 \chi -16\kappa \right)} \bigg({\bar\phi}^2 f''({\bar\phi}) +  \sigma_\mathrm{FH}^{} \left(4 -\tfrac{1}{2 \kappa} {\bar\phi}^2  \chi\right) 
    - 2 \sqrt{{\bar\phi}^2 f''({\bar\phi}) \left( \tfrac{4 \kappa f''({\bar\phi}) }{ \chi} -2 \sigma_\mathrm{FH}^{} \right) + 4 \sigma_\mathrm{FH}^2}
    \bigg)\bigg]^\frac12 ,
    \label{eqn:c0crit}
\end{align}
where $\sigma_\mathrm{FH}^{} = \sigma + f({\bar\phi})$. 
\end{widetext}
So far, we have derived the critical protein-induced curvature for the transition from a conserved Turing-type to a Cahn-Hilliard type instability, as well as the critical protein-induced curvature that defines the spinodal in the conserved Turing-type instability region. 
Next, we aim to determine the spinodal for the region with a Cahn-Hilliard-type instability, given by ${k_+ =0}$. 
Again, we recover the protein induced curvature $C_0^{*}(\chi , \kappa)$ as a solution, which is only valid for ${\tfrac12 \chi \sigma + \kappa f''({\phi_0}) \geq 0}$. 

Taken together, ${C_0^{*}(\chi , \kappa)}$, defines a surface in the parameter space (depicted as the magenta manifold in Fig.~\ref{fig:lsa}(f)), at which a sign change in the dispersion relation at the origin occurs. 
Below the magenta surface, the dispersion relation at the origin is positive, leading to a CH-type instability. 
Above the black line (${\tfrac12 \chi \sigma + \kappa f''({\phi_0}) \geq 0}$) on this surface, $C_0^{*}(\chi , \kappa)$ also satisfies the equation ${k_+ = 0}$, thus defining the boundary between a homogeneous system and a phase-separated system.
Below the black line ($ \tfrac12 \chi \sigma + \kappa f''({\phi_0}) \leq 0$) on this surface, $C_0^{*}(\chi , \kappa)$ solves the equation ${k_-=0}$, thereby defining the boundary between a CH-type and cT-type instability. 
In this case, the spinodal is given by  $C_0^{\mathrm{c}}$ and is shown as the cyan surface in Fig.~\ref{fig:lsa}(f).

\section{Comparison of Linear Stability Analyses of Dynamic Equations and Free Energy}
\label{append:comparisonLSA}

\begin{figure}[tb]
\centering
\includegraphics[width=0.95\columnwidth]{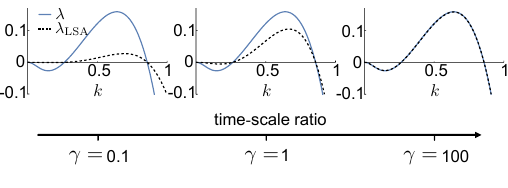}
\caption{Comparison of the dispersion relation of the full model $\lambda_\mathrm{LSA}$ (dashed line) with that of the effective free energy functional $\lambda$ (solid line). The dispersion relations agree in the limit of strong timescale separation. The ranges of unstable modes agree independently of the timescale separation, i.e., $\gamma\gg 1$. The parameters are the same as in Fig.~\ref{fig:lsa}; specifically we chose ${C_0 = 0.5}$ and vary $\gamma$.}
\label{fig:comp_lsa}
\end{figure}

We compare the dispersion relation derived from the standard linear stability analysis $\lambda_\mathrm{LSA}\left(k\right)$ with that derived from the effective free energy functional. The dispersion relation derived from the effective free energy functional, in which the height fluctuations have been integrated out, is valid if we can assume that the height field has already equilibrated on the timescale of the protein dynamics. For a positive effective surface tension this is equivalent to saying that dynamic changes in the height field have to be fast compared to the protein dynamics. The dispersion relations for different timescale ratios are shown in Fig.~\ref{fig:comp_lsa}. We make two important observations: The effective dispersion relation $\lambda(k)$ follows from the full dispersion relation $\lambda_\mathrm{LSA}(k)$ in the limit of strong timescale separation. Second, the roots of the dispersion relations agree independently of the timescale separation, and, thus, also the phase boundaries $C_0^c$ and $C_0^*$, derived in Appendix~\ref{append:LSA_FE}, don't depend on the timescales. 

While the detailed form of the dispersion relations is distinct from the expressions obtained by analyzing the effective free energy (see Appendix~\ref{append:comparisonLSA}), the zeros (marginal modes) of the dispersion relations, Eqs.~\eqref{eq:kmnius} and~\eqref{eq:kplus}, agree.
Thus, the bifurcation diagrams that differentiate between regions where the homogeneous state is stable and various types of lateral instabilities are the same for the full LSA and the LSA based on the effective free energy functional.
This reflects that the underlying system is thermodynamic, implying that the final steady state should be independent of specific timescales. By integrating out the height field dynamics to quadratic order, one effectively assumes a factorization of the contributions within the free energy. This factorization directly translates into the resulting effective dispersion relation, thereby preserving its characteristic roots.

\section{Distance measure}
\label{append:distmeasure}

\begin{figure}[!tb]
\centering
\includegraphics[width=\linewidth]{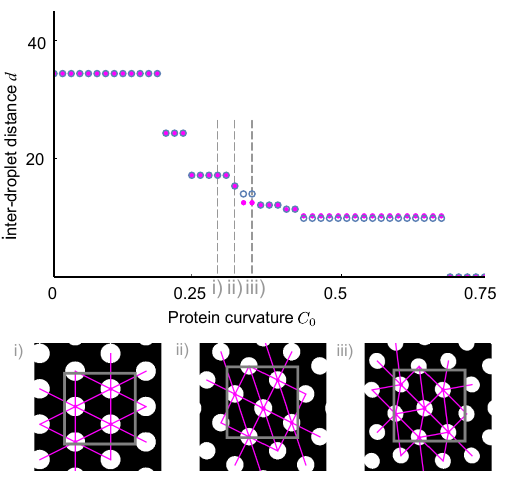}
\caption{ Droplet distances in numerical simulations as determined from center-to-center distances (pink dots) and as derived from the droplet number (blue circles) as a function of the protein induced curvature $C_0$. 
The measured center-to-center distance of the droplets is obtained by minimizing over the distances to all neighbors that can be reached without intersecting another droplet in the finite domain (gray box). Bottom panels i)-iii) show three simulation snapshots for different values of $C_0$ as indicated by dashed lines. The center-to-center distances to the neighboring droplets are indicated by pink lines in the snapshots.
Geometric frustration can be observed in panels ii) and iii).
}
\label{fig:SI_distancemeasure}
\end{figure}

In this section, we explain the method used to define an inter-droplet distance on the basis of the integer number of droplets $N$ and compare it with the inter-droplet distance as measured by minimizing over mutual center-to-center distances of neighboring droplets.
Due to geometric frustration, deviations from the regular lattice may occur, leading to slight variations in the inter-droplet distances.

We define the number-derived inter-droplet distance as the characteristic length scale of the pattern 
\begin{align}
    d_N
    =
    \sqrt{\frac{A}{N}}
    \, ,
\end{align}
where $N$ is the number of droplets in a domain with area $A$. In Fig.~\ref{fig:SI_distancemeasure}, we compare the number-derived inter-droplet distance (blue circles) to the measured distances from simulation snapshots (magenta dots). 
The latter is determined by minimizing over the distances to all the droplets that can be reached without intersecting another droplet. 
We use periodic boundary conditions to calculate the distance for droplets close to the boundary of the system. Since the simulations are performed in a finite domain (gray box), the pattern can be geometrically frustrated and deviate from the regular lattice for some parameter values, as shown in panels ii) and iii) in Fig.~\ref{fig:SI_distancemeasure}. 
The number-derived droplet distance agrees very well with the minimal distance from the simulation snapshots. 

\clearpage


\begin{thebibliography}{78}%
\makeatletter
\providecommand \@ifxundefined [1]{%
 \@ifx{#1\undefined}
}%
\providecommand \@ifnum [1]{%
 \ifnum #1\expandafter \@firstoftwo
 \else \expandafter \@secondoftwo
 \fi
}%
\providecommand \@ifx [1]{%
 \ifx #1\expandafter \@firstoftwo
 \else \expandafter \@secondoftwo
 \fi
}%
\providecommand \natexlab [1]{#1}%
\providecommand \enquote  [1]{``#1''}%
\providecommand \bibnamefont  [1]{#1}%
\providecommand \bibfnamefont [1]{#1}%
\providecommand \citenamefont [1]{#1}%
\providecommand \href@noop [0]{\@secondoftwo}%
\providecommand \href [0]{\begingroup \@sanitize@url \@href}%
\providecommand \@href[1]{\@@startlink{#1}\@@href}%
\providecommand \@@href[1]{\endgroup#1\@@endlink}%
\providecommand \@sanitize@url [0]{\catcode `\\12\catcode `\$12\catcode
  `\&12\catcode `\#12\catcode `\^12\catcode `\_12\catcode `\%12\relax}%
\providecommand \@@startlink[1]{}%
\providecommand \@@endlink[0]{}%
\providecommand \url  [0]{\begingroup\@sanitize@url \@url }%
\providecommand \@url [1]{\endgroup\@href {#1}{\urlprefix }}%
\providecommand \urlprefix  [0]{URL }%
\providecommand \Eprint [0]{\href }%
\providecommand \doibase [0]{https://doi.org/}%
\providecommand \selectlanguage [0]{\@gobble}%
\providecommand \bibinfo  [0]{\@secondoftwo}%
\providecommand \bibfield  [0]{\@secondoftwo}%
\providecommand \translation [1]{[#1]}%
\providecommand \BibitemOpen [0]{}%
\providecommand \bibitemStop [0]{}%
\providecommand \bibitemNoStop [0]{.\EOS\space}%
\providecommand \EOS [0]{\spacefactor3000\relax}%
\providecommand \BibitemShut  [1]{\csname bibitem#1\endcsname}%
\let\auto@bib@innerbib\@empty
\bibitem [{\citenamefont {Brangwynne}\ \emph {et~al.}(2009)\citenamefont
  {Brangwynne}, \citenamefont {Eckmann}, \citenamefont {Courson}, \citenamefont
  {Rybarska}, \citenamefont {Hoege}, \citenamefont {Gharakhani}, \citenamefont
  {Jülicher},\ and\ \citenamefont {Hyman}}]{Brangwynne2009Germline}%
  \BibitemOpen
  \bibfield  {author} {\bibinfo {author} {\bibfnamefont {C.~P.}\ \bibnamefont
  {Brangwynne}}, \bibinfo {author} {\bibfnamefont {C.~R.}\ \bibnamefont
  {Eckmann}}, \bibinfo {author} {\bibfnamefont {D.~S.}\ \bibnamefont
  {Courson}}, \bibinfo {author} {\bibfnamefont {A.}~\bibnamefont {Rybarska}},
  \bibinfo {author} {\bibfnamefont {C.}~\bibnamefont {Hoege}}, \bibinfo
  {author} {\bibfnamefont {J.}~\bibnamefont {Gharakhani}}, \bibinfo {author}
  {\bibfnamefont {F.}~\bibnamefont {Jülicher}},\ and\ \bibinfo {author}
  {\bibfnamefont {A.~A.}\ \bibnamefont {Hyman}},\ }\bibfield  {title} {\bibinfo
  {title} {{Germline P Granules Are Liquid Droplets That Localize by Controlled
  Dissolution/Condensation}},\ }\href {https://doi.org/10.1126/science.1172046}
  {\bibfield  {journal} {\bibinfo  {journal} {Science}\ }\textbf {\bibinfo
  {volume} {324}},\ \bibinfo {pages} {1729} (\bibinfo {year}
  {2009})}\BibitemShut {NoStop}%
\bibitem [{\citenamefont {Hyman}\ \emph {et~al.}(2014)\citenamefont {Hyman},
  \citenamefont {Weber},\ and\ \citenamefont {Jülicher}}]{Hyman:2014}%
  \BibitemOpen
  \bibfield  {author} {\bibinfo {author} {\bibfnamefont {A.~A.}\ \bibnamefont
  {Hyman}}, \bibinfo {author} {\bibfnamefont {C.~A.}\ \bibnamefont {Weber}},\
  and\ \bibinfo {author} {\bibfnamefont {F.}~\bibnamefont {Jülicher}},\
  }\bibfield  {title} {\bibinfo {title} {{Liquid-Liquid Phase Separation in
  Biology}},\ }\href {https://doi.org/10.1146/annurev-cellbio-100913-013325}
  {\bibfield  {journal} {\bibinfo  {journal} {Annu Rev Cell Dev Bi}\ }\textbf
  {\bibinfo {volume} {30}},\ \bibinfo {pages} {39} (\bibinfo {year}
  {2014})}\BibitemShut {NoStop}%
\bibitem [{\citenamefont {Doi}(2013)}]{Doi2013Soft}%
  \BibitemOpen
  \bibfield  {author} {\bibinfo {author} {\bibfnamefont {M.}~\bibnamefont
  {Doi}},\ }\href {https://doi.org/10.1093/acprof:oso/9780199652952.001.0001}
  {\emph {\bibinfo {title} {{Soft Matter Physics}}}}\ (\bibinfo  {publisher}
  {Oxford University Press},\ \bibinfo {year} {2013})\BibitemShut {NoStop}%
\bibitem [{\citenamefont {John}\ and\ \citenamefont
  {Bär}(2005)}]{John2005Traveling}%
  \BibitemOpen
  \bibfield  {author} {\bibinfo {author} {\bibfnamefont {K.}~\bibnamefont
  {John}}\ and\ \bibinfo {author} {\bibfnamefont {M.}~\bibnamefont {Bär}},\
  }\bibfield  {title} {\bibinfo {title} {{Travelling lipid domains in a dynamic
  model for protein-induced pattern formation in biomembranes}},\ }\href
  {https://doi.org/10.1088/1478-3975/2/2/005} {\bibfield  {journal} {\bibinfo
  {journal} {Phys Biol}\ }\textbf {\bibinfo {volume} {2}},\ \bibinfo {pages}
  {123} (\bibinfo {year} {2005})}\BibitemShut {NoStop}%
\bibitem [{\citenamefont {Veatch}\ and\ \citenamefont
  {Keller}(2003)}]{Veatch2003Separation}%
  \BibitemOpen
  \bibfield  {author} {\bibinfo {author} {\bibfnamefont {S.~L.}\ \bibnamefont
  {Veatch}}\ and\ \bibinfo {author} {\bibfnamefont {S.~L.}\ \bibnamefont
  {Keller}},\ }\bibfield  {title} {\bibinfo {title} {{Separation of Liquid
  Phases in Giant Vesicles of Ternary Mixtures of Phospholipids and
  Cholesterol}},\ }\href {https://doi.org/10.1016/s0006-3495(03)74726-2}
  {\bibfield  {journal} {\bibinfo  {journal} {Biophys J}\ }\textbf {\bibinfo
  {volume} {85}},\ \bibinfo {pages} {3074} (\bibinfo {year}
  {2003})}\BibitemShut {NoStop}%
\bibitem [{\citenamefont {Groves}(2007)}]{Groves2007Bending}%
  \BibitemOpen
  \bibfield  {author} {\bibinfo {author} {\bibfnamefont {J.~T.}\ \bibnamefont
  {Groves}},\ }\bibfield  {title} {\bibinfo {title} {{Bending Mechanics and
  Molecular Organization in Biological Membranes}},\ }\href
  {https://doi.org/10.1146/annurev.physchem.56.092503.141216} {\bibfield
  {journal} {\bibinfo  {journal} {Phys Chem}\ }\textbf {\bibinfo {volume}
  {58}},\ \bibinfo {pages} {697} (\bibinfo {year} {2007})}\BibitemShut
  {NoStop}%
\bibitem [{\citenamefont {Alonso}\ and\ \citenamefont
  {Bär}(2010)}]{Alonso2010Phase}%
  \BibitemOpen
  \bibfield  {author} {\bibinfo {author} {\bibfnamefont {S.}~\bibnamefont
  {Alonso}}\ and\ \bibinfo {author} {\bibfnamefont {M.}~\bibnamefont {Bär}},\
  }\bibfield  {title} {\bibinfo {title} {{Phase separation and bistability in a
  three-dimensional model for protein domain formation at biomembranes}},\
  }\href {https://doi.org/10.1088/1478-3975/7/4/046012} {\bibfield  {journal}
  {\bibinfo  {journal} {Phys Biol}\ }\textbf {\bibinfo {volume} {7}},\ \bibinfo
  {pages} {046012} (\bibinfo {year} {2010})}\BibitemShut {NoStop}%
\bibitem [{\citenamefont {Simons}\ and\ \citenamefont
  {Toomre}(2000)}]{Simons2000Lipid}%
  \BibitemOpen
  \bibfield  {author} {\bibinfo {author} {\bibfnamefont {K.}~\bibnamefont
  {Simons}}\ and\ \bibinfo {author} {\bibfnamefont {D.}~\bibnamefont
  {Toomre}},\ }\bibfield  {title} {\bibinfo {title} {{Lipid rafts and signal
  transduction}},\ }\href {https://doi.org/10.1038/35036052} {\bibfield
  {journal} {\bibinfo  {journal} {Nat Rev Mol Cell Biol}\ }\textbf {\bibinfo
  {volume} {1}},\ \bibinfo {pages} {31} (\bibinfo {year} {2000})}\BibitemShut
  {NoStop}%
\bibitem [{\citenamefont {Postma}\ \emph {et~al.}(2004)\citenamefont {Postma},
  \citenamefont {Roelofs}, \citenamefont {Goedhart}, \citenamefont {Loovers},
  \citenamefont {Visser},\ and\ \citenamefont
  {Haastert}}]{Postma2004Sensitization}%
  \BibitemOpen
  \bibfield  {author} {\bibinfo {author} {\bibfnamefont {M.}~\bibnamefont
  {Postma}}, \bibinfo {author} {\bibfnamefont {J.}~\bibnamefont {Roelofs}},
  \bibinfo {author} {\bibfnamefont {J.}~\bibnamefont {Goedhart}}, \bibinfo
  {author} {\bibfnamefont {H.~M.}\ \bibnamefont {Loovers}}, \bibinfo {author}
  {\bibfnamefont {A.~J. W.~G.}\ \bibnamefont {Visser}},\ and\ \bibinfo {author}
  {\bibfnamefont {P.~J. M.~V.}\ \bibnamefont {Haastert}},\ }\bibfield  {title}
  {\bibinfo {title} {{Sensitization of Dictyostelium chemotaxis by
  phosphoinositide-3-kinase-mediated self-organizing signalling patches}},\
  }\href {https://doi.org/10.1242/jcs.01143} {\bibfield  {journal} {\bibinfo
  {journal} {J Cell Sci}\ }\textbf {\bibinfo {volume} {117}},\ \bibinfo {pages}
  {2925} (\bibinfo {year} {2004})}\BibitemShut {NoStop}%
\bibitem [{\citenamefont {Shelby}\ \emph {et~al.}(2023)\citenamefont {Shelby},
  \citenamefont {Castello-Serrano}, \citenamefont {Wisser}, \citenamefont
  {Levental},\ and\ \citenamefont {Veatch}}]{Shelby2023Membrane}%
  \BibitemOpen
  \bibfield  {author} {\bibinfo {author} {\bibfnamefont {S.~A.}\ \bibnamefont
  {Shelby}}, \bibinfo {author} {\bibfnamefont {I.}~\bibnamefont
  {Castello-Serrano}}, \bibinfo {author} {\bibfnamefont {K.~C.}\ \bibnamefont
  {Wisser}}, \bibinfo {author} {\bibfnamefont {I.}~\bibnamefont {Levental}},\
  and\ \bibinfo {author} {\bibfnamefont {S.~L.}\ \bibnamefont {Veatch}},\
  }\bibfield  {title} {\bibinfo {title} {{Membrane phase separation drives
  responsive assembly of receptor signaling domains}},\ }\href
  {https://doi.org/10.1038/s41589-023-01268-8} {\bibfield  {journal} {\bibinfo
  {journal} {Nat Chem Biol}\ }\textbf {\bibinfo {volume} {19}},\ \bibinfo
  {pages} {750} (\bibinfo {year} {2023})}\BibitemShut {NoStop}%
\bibitem [{\citenamefont {Ikonen}(2001)}]{Ikonen2001Roles}%
  \BibitemOpen
  \bibfield  {author} {\bibinfo {author} {\bibfnamefont {E.}~\bibnamefont
  {Ikonen}},\ }\bibfield  {title} {\bibinfo {title} {{Roles of lipid rafts in
  membrane transport}},\ }\href {https://doi.org/10.1016/s0955-0674(00)00238-6}
  {\bibfield  {journal} {\bibinfo  {journal} {Curr Opin Cell Biol}\ }\textbf
  {\bibinfo {volume} {13}},\ \bibinfo {pages} {470} (\bibinfo {year}
  {2001})}\BibitemShut {NoStop}%
\bibitem [{\citenamefont {Leibler}(1986)}]{Leibler:1986}%
  \BibitemOpen
  \bibfield  {author} {\bibinfo {author} {\bibfnamefont {S.}~\bibnamefont
  {Leibler}},\ }\bibfield  {title} {\bibinfo {title} {Curvature instability in
  membranes},\ }\href {https://doi.org/10.1051/jphys:01986004703050700}
  {\bibfield  {journal} {\bibinfo  {journal} {J Phys-Paris}\ }\textbf {\bibinfo
  {volume} {47}},\ \bibinfo {pages} {507} (\bibinfo {year} {1986})}\BibitemShut
  {NoStop}%
\bibitem [{\citenamefont {Girard}\ \emph {et~al.}(2005)\citenamefont {Girard},
  \citenamefont {Prost},\ and\ \citenamefont {Bassereau}}]{Girad2005Passive}%
  \BibitemOpen
  \bibfield  {author} {\bibinfo {author} {\bibfnamefont {P.}~\bibnamefont
  {Girard}}, \bibinfo {author} {\bibfnamefont {J.}~\bibnamefont {Prost}},\ and\
  \bibinfo {author} {\bibfnamefont {P.}~\bibnamefont {Bassereau}},\ }\bibfield
  {title} {\bibinfo {title} {Passive or active fluctuations in membranes
  containing proteins},\ }\href {https://doi.org/10.1103/PhysRevLett.94.088102}
  {\bibfield  {journal} {\bibinfo  {journal} {Phys Rev Lett}\ }\textbf
  {\bibinfo {volume} {94}},\ \bibinfo {pages} {088102} (\bibinfo {year}
  {2005})}\BibitemShut {NoStop}%
\bibitem [{\citenamefont {Farsad}\ and\ \citenamefont
  {Camilli}(2003)}]{Farsad:2003}%
  \BibitemOpen
  \bibfield  {author} {\bibinfo {author} {\bibfnamefont {K.}~\bibnamefont
  {Farsad}}\ and\ \bibinfo {author} {\bibfnamefont {P.~D.}\ \bibnamefont
  {Camilli}},\ }\bibfield  {title} {\bibinfo {title} {{Mechanisms of membrane
  deformation}},\ }\href {https://doi.org/10.1016/s0955-0674(03)00073-5}
  {\bibfield  {journal} {\bibinfo  {journal} {Curr Opin Cell Biol}\ }\textbf
  {\bibinfo {volume} {15}},\ \bibinfo {pages} {372} (\bibinfo {year}
  {2003})}\BibitemShut {NoStop}%
\bibitem [{\citenamefont {Pr{\'e}vost}\ \emph {et~al.}(2015)\citenamefont
  {Pr{\'e}vost}, \citenamefont {Zhao}, \citenamefont {Manzi}, \citenamefont
  {Lemichez}, \citenamefont {Lappalainen}, \citenamefont {{Callan-Jones}},\
  and\ \citenamefont {Bassereau}}]{prevost2015irsp53}%
  \BibitemOpen
  \bibfield  {author} {\bibinfo {author} {\bibfnamefont {C.}~\bibnamefont
  {Pr{\'e}vost}}, \bibinfo {author} {\bibfnamefont {H.}~\bibnamefont {Zhao}},
  \bibinfo {author} {\bibfnamefont {J.}~\bibnamefont {Manzi}}, \bibinfo
  {author} {\bibfnamefont {E.}~\bibnamefont {Lemichez}}, \bibinfo {author}
  {\bibfnamefont {P.}~\bibnamefont {Lappalainen}}, \bibinfo {author}
  {\bibfnamefont {A.}~\bibnamefont {{Callan-Jones}}},\ and\ \bibinfo {author}
  {\bibfnamefont {P.}~\bibnamefont {Bassereau}},\ }\bibfield  {title} {\bibinfo
  {title} {{{IRSp53}} senses negative membrane curvature and phase separates
  along membrane tubules},\ }\href {https://doi.org/10.1038/ncomms9529}
  {\bibfield  {journal} {\bibinfo  {journal} {Nat Commun}\ }\textbf {\bibinfo
  {volume} {6}},\ \bibinfo {pages} {8529} (\bibinfo {year} {2015})}\BibitemShut
  {NoStop}%
\bibitem [{\citenamefont {Goychuk}\ and\ \citenamefont
  {Frey}(2019)}]{Goychuk:2019}%
  \BibitemOpen
  \bibfield  {author} {\bibinfo {author} {\bibfnamefont {A.}~\bibnamefont
  {Goychuk}}\ and\ \bibinfo {author} {\bibfnamefont {E.}~\bibnamefont {Frey}},\
  }\bibfield  {title} {\bibinfo {title} {{Protein Recruitment through Indirect
  Mechanochemical Interactions}},\ }\href
  {https://doi.org/10.1103/physrevlett.123.178101} {\bibfield  {journal}
  {\bibinfo  {journal} {Phys Rev Lett}\ }\textbf {\bibinfo {volume} {123}},\
  \bibinfo {pages} {178101} (\bibinfo {year} {2019})}\BibitemShut {NoStop}%
\bibitem [{\citenamefont {Simunovic}\ \emph {et~al.}(2016)\citenamefont
  {Simunovic}, \citenamefont {Evergren}, \citenamefont {Golushko},
  \citenamefont {Pr{\'e}vost}, \citenamefont {Renard}, \citenamefont
  {Johannes}, \citenamefont {McMahon}, \citenamefont {Lorman}, \citenamefont
  {Voth},\ and\ \citenamefont {Bassereau}}]{simunovic2016how}%
  \BibitemOpen
  \bibfield  {author} {\bibinfo {author} {\bibfnamefont {M.}~\bibnamefont
  {Simunovic}}, \bibinfo {author} {\bibfnamefont {E.}~\bibnamefont {Evergren}},
  \bibinfo {author} {\bibfnamefont {I.}~\bibnamefont {Golushko}}, \bibinfo
  {author} {\bibfnamefont {C.}~\bibnamefont {Pr{\'e}vost}}, \bibinfo {author}
  {\bibfnamefont {H.-F.}\ \bibnamefont {Renard}}, \bibinfo {author}
  {\bibfnamefont {L.}~\bibnamefont {Johannes}}, \bibinfo {author}
  {\bibfnamefont {H.~T.}\ \bibnamefont {McMahon}}, \bibinfo {author}
  {\bibfnamefont {V.}~\bibnamefont {Lorman}}, \bibinfo {author} {\bibfnamefont
  {G.~A.}\ \bibnamefont {Voth}},\ and\ \bibinfo {author} {\bibfnamefont
  {P.}~\bibnamefont {Bassereau}},\ }\bibfield  {title} {\bibinfo {title} {How
  curvature-generating proteins build scaffolds on membrane nanotubes},\ }\href
  {https://doi.org/10.1073/pnas.1606943113} {\bibfield  {journal} {\bibinfo
  {journal} {P Natl A Sci USA}\ }\textbf {\bibinfo {volume} {113}},\ \bibinfo
  {pages} {11226} (\bibinfo {year} {2016})}\BibitemShut {NoStop}%
\bibitem [{\citenamefont {Heinrich}\ \emph
  {et~al.}(2010{\natexlab{a}})\citenamefont {Heinrich}, \citenamefont {Tian},
  \citenamefont {Esposito},\ and\ \citenamefont {Baumgart}}]{Heinrich:2010}%
  \BibitemOpen
  \bibfield  {author} {\bibinfo {author} {\bibfnamefont {M.}~\bibnamefont
  {Heinrich}}, \bibinfo {author} {\bibfnamefont {A.}~\bibnamefont {Tian}},
  \bibinfo {author} {\bibfnamefont {C.}~\bibnamefont {Esposito}},\ and\
  \bibinfo {author} {\bibfnamefont {T.}~\bibnamefont {Baumgart}},\ }\bibfield
  {title} {\bibinfo {title} {{Dynamic sorting of lipids and proteins in
  membrane tubes with a moving phase boundary}},\ }\href
  {https://doi.org/10.1073/pnas.0913997107} {\bibfield  {journal} {\bibinfo
  {journal} {P Natl A Sci USA}\ }\textbf {\bibinfo {volume} {107}},\ \bibinfo
  {pages} {7208} (\bibinfo {year} {2010}{\natexlab{a}})}\BibitemShut {NoStop}%
\bibitem [{\citenamefont {Parthasarathy}\ \emph {et~al.}(2006)\citenamefont
  {Parthasarathy}, \citenamefont {Yu},\ and\ \citenamefont
  {Groves}}]{Parthasarathy2006Curvature}%
  \BibitemOpen
  \bibfield  {author} {\bibinfo {author} {\bibfnamefont {R.}~\bibnamefont
  {Parthasarathy}}, \bibinfo {author} {\bibfnamefont {C.-h.}\ \bibnamefont
  {Yu}},\ and\ \bibinfo {author} {\bibfnamefont {J.~T.}\ \bibnamefont
  {Groves}},\ }\bibfield  {title} {\bibinfo {title} {{Curvature-Modulated Phase
  Separation in Lipid Bilayer Membranes}},\ }\href
  {https://doi.org/10.1021/la060390o} {\bibfield  {journal} {\bibinfo
  {journal} {Langmuir}\ }\textbf {\bibinfo {volume} {22}},\ \bibinfo {pages}
  {5095} (\bibinfo {year} {2006})}\BibitemShut {NoStop}%
\bibitem [{\citenamefont {Heberle}\ and\ \citenamefont
  {Feigenson}(2011)}]{Heberle2011Phase}%
  \BibitemOpen
  \bibfield  {author} {\bibinfo {author} {\bibfnamefont {F.~A.}\ \bibnamefont
  {Heberle}}\ and\ \bibinfo {author} {\bibfnamefont {G.~W.}\ \bibnamefont
  {Feigenson}},\ }\bibfield  {title} {\bibinfo {title} {{Phase Separation in
  Lipid Membranes}},\ }\href {https://doi.org/10.1101/cshperspect.a004630}
  {\bibfield  {journal} {\bibinfo  {journal} {CSH Perspect Biol}\ }\textbf
  {\bibinfo {volume} {3}},\ \bibinfo {pages} {a004630} (\bibinfo {year}
  {2011})}\BibitemShut {NoStop}%
\bibitem [{\citenamefont {Vandin}\ \emph {et~al.}(2016)\citenamefont {Vandin},
  \citenamefont {Marenduzzo}, \citenamefont {Goryachev},\ and\ \citenamefont
  {Orlandini}}]{Vandin2016Curvature}%
  \BibitemOpen
  \bibfield  {author} {\bibinfo {author} {\bibfnamefont {G.}~\bibnamefont
  {Vandin}}, \bibinfo {author} {\bibfnamefont {D.}~\bibnamefont {Marenduzzo}},
  \bibinfo {author} {\bibfnamefont {A.~B.}\ \bibnamefont {Goryachev}},\ and\
  \bibinfo {author} {\bibfnamefont {E.}~\bibnamefont {Orlandini}},\ }\bibfield
  {title} {\bibinfo {title} {Curvature-driven positioning of turing patterns in
  phase-separating curved membranes},\ }\href
  {https://doi.org/10.1039/C6SM00340K} {\bibfield  {journal} {\bibinfo
  {journal} {Soft Matter}\ }\textbf {\bibinfo {volume} {12}},\ \bibinfo {pages}
  {3888} (\bibinfo {year} {2016})}\BibitemShut {NoStop}%
\bibitem [{\citenamefont {Callan-Jones}\ \emph {et~al.}(2011)\citenamefont
  {Callan-Jones}, \citenamefont {Sorre},\ and\ \citenamefont
  {Bassereau}}]{Callan-Jones2011Curvature}%
  \BibitemOpen
  \bibfield  {author} {\bibinfo {author} {\bibfnamefont {A.}~\bibnamefont
  {Callan-Jones}}, \bibinfo {author} {\bibfnamefont {B.}~\bibnamefont
  {Sorre}},\ and\ \bibinfo {author} {\bibfnamefont {P.}~\bibnamefont
  {Bassereau}},\ }\bibfield  {title} {\bibinfo {title} {{Curvature-Driven Lipid
  Sorting in Biomembranes}},\ }\href
  {https://doi.org/10.1101/cshperspect.a004648} {\bibfield  {journal} {\bibinfo
   {journal} {CSH Perspect Biol}\ }\textbf {\bibinfo {volume} {3}},\ \bibinfo
  {pages} {a004648} (\bibinfo {year} {2011})}\BibitemShut {NoStop}%
\bibitem [{\citenamefont {Roux}\ \emph {et~al.}(2005)\citenamefont {Roux},
  \citenamefont {Cuvelier}, \citenamefont {Nassoy}, \citenamefont {Prost},
  \citenamefont {Bassereau},\ and\ \citenamefont {Goud}}]{Roux2005Role}%
  \BibitemOpen
  \bibfield  {author} {\bibinfo {author} {\bibfnamefont {A.}~\bibnamefont
  {Roux}}, \bibinfo {author} {\bibfnamefont {D.}~\bibnamefont {Cuvelier}},
  \bibinfo {author} {\bibfnamefont {P.}~\bibnamefont {Nassoy}}, \bibinfo
  {author} {\bibfnamefont {J.}~\bibnamefont {Prost}}, \bibinfo {author}
  {\bibfnamefont {P.}~\bibnamefont {Bassereau}},\ and\ \bibinfo {author}
  {\bibfnamefont {B.}~\bibnamefont {Goud}},\ }\bibfield  {title} {\bibinfo
  {title} {{Role of curvature and phase transition in lipid sorting and fission
  of membrane tubules}},\ }\href {https://doi.org/10.1038/sj.emboj.7600631}
  {\bibfield  {journal} {\bibinfo  {journal} {EMBO J}\ }\textbf {\bibinfo
  {volume} {24}},\ \bibinfo {pages} {1537} (\bibinfo {year}
  {2005})}\BibitemShut {NoStop}%
\bibitem [{\citenamefont {Semrau}\ and\ \citenamefont
  {Schmidt}(2009)}]{Semrau2009Membrane}%
  \BibitemOpen
  \bibfield  {author} {\bibinfo {author} {\bibfnamefont {S.}~\bibnamefont
  {Semrau}}\ and\ \bibinfo {author} {\bibfnamefont {T.}~\bibnamefont
  {Schmidt}},\ }\bibfield  {title} {\bibinfo {title} {Membrane heterogeneity
  – from lipid domains to curvature effects},\ }\href
  {https://doi.org/10.1039/B901587F} {\bibfield  {journal} {\bibinfo  {journal}
  {Soft Matter}\ }\textbf {\bibinfo {volume} {5}},\ \bibinfo {pages} {3174}
  (\bibinfo {year} {2009})}\BibitemShut {NoStop}%
\bibitem [{\citenamefont {Idema}\ \emph {et~al.}(2010)\citenamefont {Idema},
  \citenamefont {Semrau}, \citenamefont {Storm},\ and\ \citenamefont
  {Schmidt}}]{Idema2010Membrane}%
  \BibitemOpen
  \bibfield  {author} {\bibinfo {author} {\bibfnamefont {T.}~\bibnamefont
  {Idema}}, \bibinfo {author} {\bibfnamefont {S.}~\bibnamefont {Semrau}},
  \bibinfo {author} {\bibfnamefont {C.}~\bibnamefont {Storm}},\ and\ \bibinfo
  {author} {\bibfnamefont {T.}~\bibnamefont {Schmidt}},\ }\bibfield  {title}
  {\bibinfo {title} {{Membrane Mediated Sorting}},\ }\href
  {https://doi.org/10.1103/physrevlett.104.198102} {\bibfield  {journal}
  {\bibinfo  {journal} {Phys Rev Lett}\ }\textbf {\bibinfo {volume} {104}},\
  \bibinfo {pages} {198102} (\bibinfo {year} {2010})}\BibitemShut {NoStop}%
\bibitem [{\citenamefont {Seifert}(1993)}]{Seifert1993Curvature}%
  \BibitemOpen
  \bibfield  {author} {\bibinfo {author} {\bibfnamefont {U.}~\bibnamefont
  {Seifert}},\ }\bibfield  {title} {\bibinfo {title} {Curvature-induced lateral
  phase segregation in two-component vesicles},\ }\href
  {https://doi.org/10.1103/PhysRevLett.70.1335} {\bibfield  {journal} {\bibinfo
   {journal} {Phys Rev Lett}\ }\textbf {\bibinfo {volume} {70}},\ \bibinfo
  {pages} {1335} (\bibinfo {year} {1993})}\BibitemShut {NoStop}%
\bibitem [{\citenamefont {Lavrentovich}\ \emph {et~al.}(2016)\citenamefont
  {Lavrentovich}, \citenamefont {Horsley}, \citenamefont {Radja}, \citenamefont
  {Sweeney},\ and\ \citenamefont {Kamien}}]{Lavrentovich2016First}%
  \BibitemOpen
  \bibfield  {author} {\bibinfo {author} {\bibfnamefont {M.~O.}\ \bibnamefont
  {Lavrentovich}}, \bibinfo {author} {\bibfnamefont {E.~M.}\ \bibnamefont
  {Horsley}}, \bibinfo {author} {\bibfnamefont {A.}~\bibnamefont {Radja}},
  \bibinfo {author} {\bibfnamefont {A.~M.}\ \bibnamefont {Sweeney}},\ and\
  \bibinfo {author} {\bibfnamefont {R.~D.}\ \bibnamefont {Kamien}},\ }\bibfield
   {title} {\bibinfo {title} {{First-order patterning transitions on a sphere
  as a route to cell morphology}},\ }\href
  {https://doi.org/10.1073/pnas.1600296113} {\bibfield  {journal} {\bibinfo
  {journal} {P Natl A Sci USA}\ }\textbf {\bibinfo {volume} {113}},\ \bibinfo
  {pages} {5189} (\bibinfo {year} {2016})}\BibitemShut {NoStop}%
\bibitem [{\citenamefont {Agudo-Canalejo}\ and\ \citenamefont
  {Golestanian}(2017)}]{Agudo-Canalejo2017Pattern}%
  \BibitemOpen
  \bibfield  {author} {\bibinfo {author} {\bibfnamefont {J.}~\bibnamefont
  {Agudo-Canalejo}}\ and\ \bibinfo {author} {\bibfnamefont {R.}~\bibnamefont
  {Golestanian}},\ }\bibfield  {title} {\bibinfo {title} {Pattern formation by
  curvature-inducing proteins on spherical membranes},\ }\href
  {https://doi.org/10.1088/1367-2630/aa983c} {\bibfield  {journal} {\bibinfo
  {journal} {New J Phys}\ }\textbf {\bibinfo {volume} {19}},\ \bibinfo {pages}
  {125013} (\bibinfo {year} {2017})}\BibitemShut {NoStop}%
\bibitem [{\citenamefont {Magi}\ and\ \citenamefont
  {Keener}(2017)}]{Magi2017Modelling}%
  \BibitemOpen
  \bibfield  {author} {\bibinfo {author} {\bibfnamefont {R.~E.}\ \bibnamefont
  {Magi}}\ and\ \bibinfo {author} {\bibfnamefont {J.~P.}\ \bibnamefont
  {Keener}},\ }\bibfield  {title} {\bibinfo {title} {Modelling a biological
  membrane as a two phase viscous fluid with curvature elasticity},\ }\href
  {https://doi.org/10.1137/15M1038141} {\bibfield  {journal} {\bibinfo
  {journal} {SIAM J Appl Math}\ }\textbf {\bibinfo {volume} {77}},\ \bibinfo
  {pages} {128} (\bibinfo {year} {2017})}\BibitemShut {NoStop}%
\bibitem [{\citenamefont {Mahapatra}\ \emph {et~al.}(2021)\citenamefont
  {Mahapatra}, \citenamefont {Saintillan},\ and\ \citenamefont
  {Rangamani}}]{Mahapatra:2021}%
  \BibitemOpen
  \bibfield  {author} {\bibinfo {author} {\bibfnamefont {A.}~\bibnamefont
  {Mahapatra}}, \bibinfo {author} {\bibfnamefont {D.}~\bibnamefont
  {Saintillan}},\ and\ \bibinfo {author} {\bibfnamefont {P.}~\bibnamefont
  {Rangamani}},\ }\bibfield  {title} {\bibinfo {title} {{Curvature-driven
  feedback on aggregation–diffusion of proteins in lipid bilayers}},\ }\href
  {https://doi.org/10.1039/d1sm00502b} {\bibfield  {journal} {\bibinfo
  {journal} {Soft Matter}\ }\textbf {\bibinfo {volume} {17}},\ \bibinfo {pages}
  {8373} (\bibinfo {year} {2021})}\BibitemShut {NoStop}%
\bibitem [{\citenamefont {Yu}\ and\ \citenamefont {Ko{\v
  s}mrlj}(2023)}]{yu2023pattern}%
  \BibitemOpen
  \bibfield  {author} {\bibinfo {author} {\bibfnamefont {Q.}~\bibnamefont
  {Yu}}\ and\ \bibinfo {author} {\bibfnamefont {A.}~\bibnamefont {Ko{\v
  s}mrlj}},\ }\href {https://doi.org/10.48550/arXiv.2309.05160} {\bibinfo
  {title} {Pattern formation of phase-separated lipid domains in bilayer
  membranes}} (\bibinfo {year} {2023}),\ \Eprint
  {https://arxiv.org/abs/2309.05160} {arxiv:2309.05160} \BibitemShut {NoStop}%
\bibitem [{\citenamefont {Flory}(1941)}]{Flory:1941}%
  \BibitemOpen
  \bibfield  {author} {\bibinfo {author} {\bibfnamefont {P.~J.}\ \bibnamefont
  {Flory}},\ }\bibfield  {title} {\bibinfo {title} {{Thermodynamics of High
  Polymer Solutions}},\ }\href {https://doi.org/10.1063/1.1750971} {\bibfield
  {journal} {\bibinfo  {journal} {J Chem Phys}\ }\textbf {\bibinfo {volume}
  {9}},\ \bibinfo {pages} {660–660} (\bibinfo {year} {1941})}\BibitemShut
  {NoStop}%
\bibitem [{\citenamefont {Huggins}(1941)}]{Huggins:1941}%
  \BibitemOpen
  \bibfield  {author} {\bibinfo {author} {\bibfnamefont {M.~L.}\ \bibnamefont
  {Huggins}},\ }\bibfield  {title} {\bibinfo {title} {{Solutions of Long Chain
  Compounds}},\ }\href {https://doi.org/10.1063/1.1750930} {\bibfield
  {journal} {\bibinfo  {journal} {J Chem Phys}\ }\textbf {\bibinfo {volume}
  {9}},\ \bibinfo {pages} {440–440} (\bibinfo {year} {1941})}\BibitemShut
  {NoStop}%
\bibitem [{\citenamefont {Canham}(1970)}]{Canham:1970}%
  \BibitemOpen
  \bibfield  {author} {\bibinfo {author} {\bibfnamefont {P.}~\bibnamefont
  {Canham}},\ }\bibfield  {title} {\bibinfo {title} {{The minimum energy of
  bending as a possible explanation of the biconcave shape of the human red
  blood cell}},\ }\href {https://doi.org/10.1016/s0022-5193(70)80032-7}
  {\bibfield  {journal} {\bibinfo  {journal} {J Theor Biol}\ }\textbf {\bibinfo
  {volume} {26}},\ \bibinfo {pages} {61–81} (\bibinfo {year}
  {1970})}\BibitemShut {NoStop}%
\bibitem [{\citenamefont {Helfrich}(1973)}]{Helfrich:1973}%
  \BibitemOpen
  \bibfield  {author} {\bibinfo {author} {\bibfnamefont {W.}~\bibnamefont
  {Helfrich}},\ }\bibfield  {title} {\bibinfo {title} {{Elastic Properties of
  Lipid Bilayers: Theory and Possible Experiments}},\ }\href
  {https://doi.org/10.1515/znc-1973-11-1209} {\bibfield  {journal} {\bibinfo
  {journal} {Z Naturforsch C}\ }\textbf {\bibinfo {volume} {28}},\ \bibinfo
  {pages} {693–703} (\bibinfo {year} {1973})}\BibitemShut {NoStop}%
\bibitem [{\citenamefont {Peter}\ \emph {et~al.}(2004)\citenamefont {Peter},
  \citenamefont {Kent}, \citenamefont {Mills}, \citenamefont {Vallis},
  \citenamefont {Butler}, \citenamefont {Evans},\ and\ \citenamefont
  {McMahon}}]{Peter:2004}%
  \BibitemOpen
  \bibfield  {author} {\bibinfo {author} {\bibfnamefont {B.~J.}\ \bibnamefont
  {Peter}}, \bibinfo {author} {\bibfnamefont {H.~M.}\ \bibnamefont {Kent}},
  \bibinfo {author} {\bibfnamefont {I.~G.}\ \bibnamefont {Mills}}, \bibinfo
  {author} {\bibfnamefont {Y.}~\bibnamefont {Vallis}}, \bibinfo {author}
  {\bibfnamefont {P.~J.~G.}\ \bibnamefont {Butler}}, \bibinfo {author}
  {\bibfnamefont {P.~R.}\ \bibnamefont {Evans}},\ and\ \bibinfo {author}
  {\bibfnamefont {H.~T.}\ \bibnamefont {McMahon}},\ }\bibfield  {title}
  {\bibinfo {title} {Bar domains as sensors of membrane curvature: The
  amphiphysin bar structure},\ }\href {https://doi.org/10.1126/science.1092586}
  {\bibfield  {journal} {\bibinfo  {journal} {Science}\ }\textbf {\bibinfo
  {volume} {303}},\ \bibinfo {pages} {495} (\bibinfo {year}
  {2004})}\BibitemShut {NoStop}%
\bibitem [{\citenamefont {Saarikangas}\ \emph {et~al.}(2009)\citenamefont
  {Saarikangas}, \citenamefont {Zhao}, \citenamefont {Pykäläinen},
  \citenamefont {Laurinmäki}, \citenamefont {Mattila}, \citenamefont
  {Kinnunen}, \citenamefont {Butcher},\ and\ \citenamefont
  {Lappalainen}}]{Saarikangas:2009}%
  \BibitemOpen
  \bibfield  {author} {\bibinfo {author} {\bibfnamefont {J.}~\bibnamefont
  {Saarikangas}}, \bibinfo {author} {\bibfnamefont {H.}~\bibnamefont {Zhao}},
  \bibinfo {author} {\bibfnamefont {A.}~\bibnamefont {Pykäläinen}}, \bibinfo
  {author} {\bibfnamefont {P.}~\bibnamefont {Laurinmäki}}, \bibinfo {author}
  {\bibfnamefont {P.~K.}\ \bibnamefont {Mattila}}, \bibinfo {author}
  {\bibfnamefont {P.~K.}\ \bibnamefont {Kinnunen}}, \bibinfo {author}
  {\bibfnamefont {S.~J.}\ \bibnamefont {Butcher}},\ and\ \bibinfo {author}
  {\bibfnamefont {P.}~\bibnamefont {Lappalainen}},\ }\bibfield  {title}
  {\bibinfo {title} {{Molecular Mechanisms of Membrane Deformation by I-BAR
  Domain Proteins}},\ }\href {https://doi.org/10.1016/j.cub.2008.12.029}
  {\bibfield  {journal} {\bibinfo  {journal} {Curr Biol}\ }\textbf {\bibinfo
  {volume} {19}},\ \bibinfo {pages} {95} (\bibinfo {year} {2009})}\BibitemShut
  {NoStop}%
\bibitem [{\citenamefont {Heinrich}\ \emph
  {et~al.}(2010{\natexlab{b}})\citenamefont {Heinrich}, \citenamefont
  {Capraro}, \citenamefont {Tian}, \citenamefont {Isas}, \citenamefont
  {Langen},\ and\ \citenamefont {Baumgart}}]{Heinrich2:2010}%
  \BibitemOpen
  \bibfield  {author} {\bibinfo {author} {\bibfnamefont {M.~C.}\ \bibnamefont
  {Heinrich}}, \bibinfo {author} {\bibfnamefont {B.~R.}\ \bibnamefont
  {Capraro}}, \bibinfo {author} {\bibfnamefont {A.}~\bibnamefont {Tian}},
  \bibinfo {author} {\bibfnamefont {J.~M.}\ \bibnamefont {Isas}}, \bibinfo
  {author} {\bibfnamefont {R.}~\bibnamefont {Langen}},\ and\ \bibinfo {author}
  {\bibfnamefont {T.}~\bibnamefont {Baumgart}},\ }\bibfield  {title} {\bibinfo
  {title} {{Quantifying membrane curvature generation of Drosophila amphiphysin
  N-BAR domains}},\ }\href@noop {} {\bibfield  {journal} {\bibinfo  {journal}
  {The journal of physical chemistry letters}\ }\textbf {\bibinfo {volume}
  {1}},\ \bibinfo {pages} {3401} (\bibinfo {year}
  {2010}{\natexlab{b}})}\BibitemShut {NoStop}%
\bibitem [{\citenamefont {Ford}\ \emph {et~al.}(2002)\citenamefont {Ford},
  \citenamefont {Mills}, \citenamefont {Peter}, \citenamefont {Vallis},
  \citenamefont {Praefcke}, \citenamefont {Evans},\ and\ \citenamefont
  {McMahon}}]{Ford:2002}%
  \BibitemOpen
  \bibfield  {author} {\bibinfo {author} {\bibfnamefont {M.~G.~J.}\
  \bibnamefont {Ford}}, \bibinfo {author} {\bibfnamefont {I.~G.}\ \bibnamefont
  {Mills}}, \bibinfo {author} {\bibfnamefont {B.~J.}\ \bibnamefont {Peter}},
  \bibinfo {author} {\bibfnamefont {Y.}~\bibnamefont {Vallis}}, \bibinfo
  {author} {\bibfnamefont {G.~J.~K.}\ \bibnamefont {Praefcke}}, \bibinfo
  {author} {\bibfnamefont {P.~R.}\ \bibnamefont {Evans}},\ and\ \bibinfo
  {author} {\bibfnamefont {H.~T.}\ \bibnamefont {McMahon}},\ }\bibfield
  {title} {\bibinfo {title} {{Curvature of clathrin-coated pits driven by
  epsin}},\ }\href {https://doi.org/10.1038/nature01020} {\bibfield  {journal}
  {\bibinfo  {journal} {Nature}\ }\textbf {\bibinfo {volume} {419}},\ \bibinfo
  {pages} {361} (\bibinfo {year} {2002})}\BibitemShut {NoStop}%
\bibitem [{\citenamefont {Kukulski}\ \emph {et~al.}(2012)\citenamefont
  {Kukulski}, \citenamefont {Schorb}, \citenamefont {Kaksonen},\ and\
  \citenamefont {Briggs}}]{Kukulski:2012}%
  \BibitemOpen
  \bibfield  {author} {\bibinfo {author} {\bibfnamefont {W.}~\bibnamefont
  {Kukulski}}, \bibinfo {author} {\bibfnamefont {M.}~\bibnamefont {Schorb}},
  \bibinfo {author} {\bibfnamefont {M.}~\bibnamefont {Kaksonen}},\ and\
  \bibinfo {author} {\bibfnamefont {J.~A.~G.}\ \bibnamefont {Briggs}},\
  }\bibfield  {title} {\bibinfo {title} {Plasma membrane reshaping during
  endocytosis is revealed by time-resolved electron tomography},\ }\href
  {https://doi.org/10.1016/j.cell.2012.05.046} {\bibfield  {journal} {\bibinfo
  {journal} {Cell}\ }\textbf {\bibinfo {volume} {150}},\ \bibinfo {pages}
  {508—520} (\bibinfo {year} {2012})}\BibitemShut {NoStop}%
\bibitem [{\citenamefont {Litschel}\ \emph {et~al.}(2018)\citenamefont
  {Litschel}, \citenamefont {Ramm}, \citenamefont {Maas}, \citenamefont
  {Heymann},\ and\ \citenamefont {Schwille}}]{litschel2018beating}%
  \BibitemOpen
  \bibfield  {author} {\bibinfo {author} {\bibfnamefont {T.}~\bibnamefont
  {Litschel}}, \bibinfo {author} {\bibfnamefont {B.}~\bibnamefont {Ramm}},
  \bibinfo {author} {\bibfnamefont {R.}~\bibnamefont {Maas}}, \bibinfo {author}
  {\bibfnamefont {M.}~\bibnamefont {Heymann}},\ and\ \bibinfo {author}
  {\bibfnamefont {P.}~\bibnamefont {Schwille}},\ }\bibfield  {title} {\bibinfo
  {title} {Beating {{Vesicles}}: {{Encapsulated Protein Oscillations Cause
  Dynamic Membrane Deformations}}},\ }\href
  {https://doi.org/10.1002/anie.201808750} {\bibfield  {journal} {\bibinfo
  {journal} {Angew Chem Int Edit}\ }\textbf {\bibinfo {volume} {57}},\ \bibinfo
  {pages} {16286} (\bibinfo {year} {2018})}\BibitemShut {NoStop}%
\bibitem [{\citenamefont {Christ}\ \emph {et~al.}(2020)\citenamefont {Christ},
  \citenamefont {Litschel}, \citenamefont {Schwille},\ and\ \citenamefont
  {Lipowsky}}]{Christ2020Active}%
  \BibitemOpen
  \bibfield  {author} {\bibinfo {author} {\bibfnamefont {S.}~\bibnamefont
  {Christ}}, \bibinfo {author} {\bibfnamefont {T.}~\bibnamefont {Litschel}},
  \bibinfo {author} {\bibfnamefont {P.}~\bibnamefont {Schwille}},\ and\
  \bibinfo {author} {\bibfnamefont {R.}~\bibnamefont {Lipowsky}},\ }\bibfield
  {title} {\bibinfo {title} {{Active shape oscillations of giant vesicles with
  cyclic closure and opening of membrane necks}},\ }\href
  {https://doi.org/10.1039/d0sm00790k} {\bibfield  {journal} {\bibinfo
  {journal} {Soft Matter}\ }\textbf {\bibinfo {volume} {17}},\ \bibinfo {pages}
  {319} (\bibinfo {year} {2020})}\BibitemShut {NoStop}%
\bibitem [{\citenamefont {{Frohoff-H\"ulsmann, Tobias and Thiele, Uwe and
  Pismen, Len M.}}(2023)}]{Frohoff2023Non}%
  \BibitemOpen
  \bibfield  {author} {\bibinfo {author} {\bibnamefont {{Frohoff-H\"ulsmann,
  Tobias and Thiele, Uwe and Pismen, Len M.}}},\ }\bibfield  {title} {\bibinfo
  {title} {{Non-reciprocity induces resonances in a two-field Cahn-Hilliard
  model}},\ }\href {https://doi.org/10.1098/rsta.2022.0087} {\bibfield
  {journal} {\bibinfo  {journal} {Phil T R Soc A}\ }\textbf {\bibinfo {volume}
  {381}},\ \bibinfo {pages} {20220087} (\bibinfo {year} {2023})},\ \Eprint
  {https://arxiv.org/abs/2211.08320} {2211.08320} \BibitemShut {NoStop}%
\bibitem [{\citenamefont {Desai}\ and\ \citenamefont
  {Kapral}(2009)}]{Desai:2009}%
  \BibitemOpen
  \bibfield  {author} {\bibinfo {author} {\bibfnamefont {R.~C.}\ \bibnamefont
  {Desai}}\ and\ \bibinfo {author} {\bibfnamefont {R.}~\bibnamefont {Kapral}},\
  }\href@noop {} {\emph {\bibinfo {title} {Dynamics of Self-organized and
  Self-assembled Structures}}}\ (\bibinfo  {publisher} {Cambridge University
  Press},\ \bibinfo {year} {2009})\BibitemShut {NoStop}%
\bibitem [{\citenamefont {Morris}\ and\ \citenamefont
  {Homann}(2001)}]{Morris:2001}%
  \BibitemOpen
  \bibfield  {author} {\bibinfo {author} {\bibfnamefont {C.}~\bibnamefont
  {Morris}}\ and\ \bibinfo {author} {\bibfnamefont {U.}~\bibnamefont
  {Homann}},\ }\bibfield  {title} {\bibinfo {title} {{Cell surface area
  regulation and membrane tension}},\ }\href@noop {} {\bibfield  {journal}
  {\bibinfo  {journal} {The Journal of membrane biology}\ }\textbf {\bibinfo
  {volume} {179}},\ \bibinfo {pages} {79} (\bibinfo {year} {2001})}\BibitemShut
  {NoStop}%
\bibitem [{\citenamefont {Deserno}(2014)}]{Deserno:2014}%
  \BibitemOpen
  \bibfield  {author} {\bibinfo {author} {\bibfnamefont {M.}~\bibnamefont
  {Deserno}},\ }\bibfield  {title} {\bibinfo {title} {Fluid lipid membranes:
  From differential geometry to curvature stresses},\ }\href
  {https://doi.org/10.1016/j.chemphyslip.2014.05.001} {\bibfield  {journal}
  {\bibinfo  {journal} {Chem Phys Lipids}\ }\textbf {\bibinfo {volume} {185}},\
  \bibinfo {pages} {11} (\bibinfo {year} {2014})}\BibitemShut {NoStop}%
\bibitem [{\citenamefont {G{\'o}{\'z}d{\'z}}(2006)}]{Gozdz:2006}%
  \BibitemOpen
  \bibfield  {author} {\bibinfo {author} {\bibfnamefont {W.~T.}\ \bibnamefont
  {G{\'o}{\'z}d{\'z}}},\ }\bibfield  {title} {\bibinfo {title} {{The Interface
  Width of Separated Two-Component Lipid Membranes}},\ }\href
  {https://doi.org/10.1021/jp062304z} {\bibfield  {journal} {\bibinfo
  {journal} {J Phys Chem B}\ }\textbf {\bibinfo {volume} {110}},\ \bibinfo
  {pages} {21981} (\bibinfo {year} {2006})}\BibitemShut {NoStop}%
\bibitem [{\citenamefont {Frey}\ and\ \citenamefont
  {Nelson}(1991)}]{Frey:1991}%
  \BibitemOpen
  \bibfield  {author} {\bibinfo {author} {\bibfnamefont {E.}~\bibnamefont
  {Frey}}\ and\ \bibinfo {author} {\bibfnamefont {D.~R.}\ \bibnamefont
  {Nelson}},\ }\bibfield  {title} {\bibinfo {title} {{Dynamics of flat
  membranes and flickering in red blood cells}},\ }\href
  {https://doi.org/10.1051/jp1:1991238} {\bibfield  {journal} {\bibinfo
  {journal} {J Phys I}\ }\textbf {\bibinfo {volume} {1}},\ \bibinfo {pages}
  {1715} (\bibinfo {year} {1991})}\BibitemShut {NoStop}%
\bibitem [{\citenamefont {{W\"urthner, Laeschkir and Goychuk, Andriy and Frey,
  Erwin}}(2023)}]{Wuerthner:2023}%
  \BibitemOpen
  \bibfield  {author} {\bibinfo {author} {\bibnamefont {{W\"urthner, Laeschkir
  and Goychuk, Andriy and Frey, Erwin}}},\ }\bibfield  {title} {\bibinfo
  {title} {{Geometry-induced patterns through mechanochemical coupling}},\
  }\href {https://doi.org/10.1103/physreve.108.014404} {\bibfield  {journal}
  {\bibinfo  {journal} {Phys Rev E}\ }\textbf {\bibinfo {volume} {108}},\
  \bibinfo {pages} {014404} (\bibinfo {year} {2023})}\BibitemShut {NoStop}%
\bibitem [{\citenamefont {Salbreux}\ and\ \citenamefont
  {J\"ulicher}(2017)}]{Salbreux2017Mechanics}%
  \BibitemOpen
  \bibfield  {author} {\bibinfo {author} {\bibfnamefont {G.}~\bibnamefont
  {Salbreux}}\ and\ \bibinfo {author} {\bibfnamefont {F.}~\bibnamefont
  {J\"ulicher}},\ }\bibfield  {title} {\bibinfo {title} {Mechanics of active
  surfaces},\ }\href {https://doi.org/10.1103/PhysRevE.96.032404} {\bibfield
  {journal} {\bibinfo  {journal} {Phys Rev E}\ }\textbf {\bibinfo {volume}
  {96}},\ \bibinfo {pages} {032404} (\bibinfo {year} {2017})}\BibitemShut
  {NoStop}%
\bibitem [{\citenamefont {De~Groot}\ and\ \citenamefont
  {Mazur}(2013)}]{degroot:2013}%
  \BibitemOpen
  \bibfield  {author} {\bibinfo {author} {\bibfnamefont {S.}~\bibnamefont
  {De~Groot}}\ and\ \bibinfo {author} {\bibfnamefont {P.}~\bibnamefont
  {Mazur}},\ }\href {https://books.google.de/books?id=mfFyG9jfaMYC} {\emph
  {\bibinfo {title} {Non-Equilibrium Thermodynamics}}},\ Dover Books on
  Physics\ (\bibinfo  {publisher} {Dover Publications},\ \bibinfo {year}
  {2013})\BibitemShut {NoStop}%
\bibitem [{\citenamefont {Cahn}\ and\ \citenamefont
  {Hilliard}(1958)}]{Cahn:1958}%
  \BibitemOpen
  \bibfield  {author} {\bibinfo {author} {\bibfnamefont {J.~W.}\ \bibnamefont
  {Cahn}}\ and\ \bibinfo {author} {\bibfnamefont {J.~E.}\ \bibnamefont
  {Hilliard}},\ }\bibfield  {title} {\bibinfo {title} {{Free Energy of a
  Nonuniform System. I. Interfacial Free Energy}},\ }\href
  {https://doi.org/10.1063/1.1744102} {\bibfield  {journal} {\bibinfo
  {journal} {J Chem Phys}\ }\textbf {\bibinfo {volume} {28}},\ \bibinfo {pages}
  {258} (\bibinfo {year} {1958})}\BibitemShut {NoStop}%
\bibitem [{\citenamefont {Cahn}(1961)}]{Cahn1961spinodal}%
  \BibitemOpen
  \bibfield  {author} {\bibinfo {author} {\bibfnamefont {J.~W.}\ \bibnamefont
  {Cahn}},\ }\bibfield  {title} {\bibinfo {title} {On spinodal decomposition},\
  }\href {https://doi.org/https://doi.org/10.1016/0001-6160(61)90182-1}
  {\bibfield  {journal} {\bibinfo  {journal} {Acta Metall Mater}\ }\textbf
  {\bibinfo {volume} {9}},\ \bibinfo {pages} {795} (\bibinfo {year}
  {1961})}\BibitemShut {NoStop}%
\bibitem [{\citenamefont {Bucher}\ \emph {et~al.}(2018)\citenamefont {Bucher},
  \citenamefont {Frey}, \citenamefont {Sochacki}, \citenamefont {Kummer},
  \citenamefont {Bergeest}, \citenamefont {Godinez}, \citenamefont
  {Kr{\"a}usslich}, \citenamefont {Rohr}, \citenamefont {Taraska},
  \citenamefont {Schwarz},\ and\ \citenamefont {Boulant}}]{Bucher:2018}%
  \BibitemOpen
  \bibfield  {author} {\bibinfo {author} {\bibfnamefont {D.}~\bibnamefont
  {Bucher}}, \bibinfo {author} {\bibfnamefont {F.}~\bibnamefont {Frey}},
  \bibinfo {author} {\bibfnamefont {K.~A.}\ \bibnamefont {Sochacki}}, \bibinfo
  {author} {\bibfnamefont {S.}~\bibnamefont {Kummer}}, \bibinfo {author}
  {\bibfnamefont {J.-P.}\ \bibnamefont {Bergeest}}, \bibinfo {author}
  {\bibfnamefont {W.~J.}\ \bibnamefont {Godinez}}, \bibinfo {author}
  {\bibfnamefont {H.-G.}\ \bibnamefont {Kr{\"a}usslich}}, \bibinfo {author}
  {\bibfnamefont {K.}~\bibnamefont {Rohr}}, \bibinfo {author} {\bibfnamefont
  {J.~W.}\ \bibnamefont {Taraska}}, \bibinfo {author} {\bibfnamefont {U.~S.}\
  \bibnamefont {Schwarz}},\ and\ \bibinfo {author} {\bibfnamefont
  {S.}~\bibnamefont {Boulant}},\ }\bibfield  {title} {\bibinfo {title}
  {Clathrin-adaptor ratio and membrane tension regulate the flat-to-curved
  transition of the clathrin coat during endocytosis},\ }\href
  {https://doi.org/10.1038/s41467-018-03533-0} {\bibfield  {journal} {\bibinfo
  {journal} {Nat Commun}\ }\textbf {\bibinfo {volume} {9}},\ \bibinfo {pages}
  {1109} (\bibinfo {year} {2018})}\BibitemShut {NoStop}%
\bibitem [{\citenamefont {Fowler}\ \emph {et~al.}(2016)\citenamefont {Fowler},
  \citenamefont {Hélie}, \citenamefont {Duncan}, \citenamefont {Chavent},
  \citenamefont {Koldsø},\ and\ \citenamefont {Sansom}}]{Fowler:2016}%
  \BibitemOpen
  \bibfield  {author} {\bibinfo {author} {\bibfnamefont {P.~W.}\ \bibnamefont
  {Fowler}}, \bibinfo {author} {\bibfnamefont {J.}~\bibnamefont {Hélie}},
  \bibinfo {author} {\bibfnamefont {A.}~\bibnamefont {Duncan}}, \bibinfo
  {author} {\bibfnamefont {M.}~\bibnamefont {Chavent}}, \bibinfo {author}
  {\bibfnamefont {H.}~\bibnamefont {Koldsø}},\ and\ \bibinfo {author}
  {\bibfnamefont {M.~S.~P.}\ \bibnamefont {Sansom}},\ }\bibfield  {title}
  {\bibinfo {title} {Membrane stiffness is modified by integral membrane
  proteins},\ }\href {https://doi.org/10.1039/C6SM01186A} {\bibfield  {journal}
  {\bibinfo  {journal} {Soft Matter}\ }\textbf {\bibinfo {volume} {12}},\
  \bibinfo {pages} {7792} (\bibinfo {year} {2016})}\BibitemShut {NoStop}%
\bibitem [{\citenamefont {{Quemeneur}}\ \emph {et~al.}(2014)\citenamefont
  {{Quemeneur}}, \citenamefont {{Sigurdsson}}, \citenamefont {{Renner}},
  \citenamefont {{Atzberger}}, \citenamefont {{Bassereau}},\ and\ \citenamefont
  {{Lacoste}}}]{Quemeneur:2014}%
  \BibitemOpen
  \bibfield  {author} {\bibinfo {author} {\bibfnamefont {F.}~\bibnamefont
  {{Quemeneur}}}, \bibinfo {author} {\bibfnamefont {J.~K.}\ \bibnamefont
  {{Sigurdsson}}}, \bibinfo {author} {\bibfnamefont {M.}~\bibnamefont
  {{Renner}}}, \bibinfo {author} {\bibfnamefont {P.~J.}\ \bibnamefont
  {{Atzberger}}}, \bibinfo {author} {\bibfnamefont {P.}~\bibnamefont
  {{Bassereau}}},\ and\ \bibinfo {author} {\bibfnamefont {D.}~\bibnamefont
  {{Lacoste}}},\ }\bibfield  {title} {\bibinfo {title} {{Shape matters in
  protein mobility within membranes}},\ }\href
  {https://doi.org/10.1073/pnas.1321054111} {\bibfield  {journal} {\bibinfo
  {journal} {P Natl A Sci USA}\ }\textbf {\bibinfo {volume} {111}},\ \bibinfo
  {pages} {5083} (\bibinfo {year} {2014})}\BibitemShut {NoStop}%
\bibitem [{\citenamefont {Baumgart}\ \emph {et~al.}(2003)\citenamefont
  {Baumgart}, \citenamefont {Hess},\ and\ \citenamefont
  {Webb}}]{Baumgart:2003}%
  \BibitemOpen
  \bibfield  {author} {\bibinfo {author} {\bibfnamefont {T.}~\bibnamefont
  {Baumgart}}, \bibinfo {author} {\bibfnamefont {S.~T.}\ \bibnamefont {Hess}},\
  and\ \bibinfo {author} {\bibfnamefont {W.~W.}\ \bibnamefont {Webb}},\
  }\bibfield  {title} {\bibinfo {title} {{Imaging coexisting fluid domains in
  biomembrane models coupling curvature and line tension}},\ }\href
  {https://doi.org/10.1038/nature02013} {\bibfield  {journal} {\bibinfo
  {journal} {Nature}\ }\textbf {\bibinfo {volume} {425}},\ \bibinfo {pages}
  {821–824} (\bibinfo {year} {2003})}\BibitemShut {NoStop}%
\bibitem [{\citenamefont {Milo}\ \emph {et~al.}(2016)\citenamefont {Milo},
  \citenamefont {Phillips},\ and\ \citenamefont {Orme}}]{Milo:2016}%
  \BibitemOpen
  \bibfield  {author} {\bibinfo {author} {\bibfnamefont {R.}~\bibnamefont
  {Milo}}, \bibinfo {author} {\bibfnamefont {R.}~\bibnamefont {Phillips}},\
  and\ \bibinfo {author} {\bibfnamefont {N.}~\bibnamefont {Orme}},\ }\href
  {https://books.google.de/books?id=GNw5jgEACAAJ} {\emph {\bibinfo {title}
  {Cell Biology by the Numbers}}}\ (\bibinfo  {publisher} {Garland Science,
  Taylor \& Francis Group},\ \bibinfo {year} {2016})\BibitemShut {NoStop}%
\bibitem [{\citenamefont {Zhong-Can}\ and\ \citenamefont
  {Helfrich}(1989)}]{ou-yang1989Bending}%
  \BibitemOpen
  \bibfield  {author} {\bibinfo {author} {\bibfnamefont {O.-Y.}\ \bibnamefont
  {Zhong-Can}}\ and\ \bibinfo {author} {\bibfnamefont {W.}~\bibnamefont
  {Helfrich}},\ }\bibfield  {title} {\bibinfo {title} {Bending energy of
  vesicle membranes: General expressions for the first, second, and third
  variation of the shape energy and applications to spheres and cylinders},\
  }\href {https://doi.org/10.1103/PhysRevA.39.5280} {\bibfield  {journal}
  {\bibinfo  {journal} {Phys. Rev. A}\ }\textbf {\bibinfo {volume} {39}},\
  \bibinfo {pages} {5280} (\bibinfo {year} {1989})}\BibitemShut {NoStop}%
\bibitem [{\citenamefont {Lifshitz}\ and\ \citenamefont
  {Slyozov}(1961)}]{Lifshitz:1961}%
  \BibitemOpen
  \bibfield  {author} {\bibinfo {author} {\bibfnamefont {I.}~\bibnamefont
  {Lifshitz}}\ and\ \bibinfo {author} {\bibfnamefont {V.}~\bibnamefont
  {Slyozov}},\ }\bibfield  {title} {\bibinfo {title} {{The kinetics of
  precipitation from supersaturated solid solutions}},\ }\href
  {https://doi.org/10.1016/0022-3697(61)90054-3} {\bibfield  {journal}
  {\bibinfo  {journal} {J Phys Chem Solids}\ }\textbf {\bibinfo {volume}
  {19}},\ \bibinfo {pages} {35} (\bibinfo {year} {1961})}\BibitemShut {NoStop}%
\bibitem [{\citenamefont {Wagner}(1961)}]{Wagner1961Theorie}%
  \BibitemOpen
  \bibfield  {author} {\bibinfo {author} {\bibfnamefont {C.}~\bibnamefont
  {Wagner}},\ }\bibfield  {title} {\bibinfo {title} {{Theorie der Alterung von
  Niederschlägen durch Umlösen (Ostwald-Reifung)}},\ }\href
  {https://doi.org/10.1002/bbpc.19610650704} {\bibfield  {journal} {\bibinfo
  {journal} {Z Elektrochem}\ }\textbf {\bibinfo {volume} {65}},\ \bibinfo
  {pages} {581} (\bibinfo {year} {1961})}\BibitemShut {NoStop}%
\bibitem [{\citenamefont {Cross}\ and\ \citenamefont
  {Hohenberg}(1993)}]{Cross-Hohenberg.1993}%
  \BibitemOpen
  \bibfield  {author} {\bibinfo {author} {\bibfnamefont {M.~C.}\ \bibnamefont
  {Cross}}\ and\ \bibinfo {author} {\bibfnamefont {P.~C.}\ \bibnamefont
  {Hohenberg}},\ }\bibfield  {title} {\bibinfo {title} {{Pattern formation
  outside of equilibrium}},\ }\href {https://doi.org/10.1103/revmodphys.65.851}
  {\bibfield  {journal} {\bibinfo  {journal} {Rev Mod Phys}\ }\textbf {\bibinfo
  {volume} {65}},\ \bibinfo {pages} {851} (\bibinfo {year} {1993})}\BibitemShut
  {NoStop}%
\bibitem [{\citenamefont {Frohoff-Hülsmann}\ and\ \citenamefont
  {Thiele}(2023)}]{Frohoff2023Nonreciprocal}%
  \BibitemOpen
  \bibfield  {author} {\bibinfo {author} {\bibfnamefont {T.}~\bibnamefont
  {Frohoff-Hülsmann}}\ and\ \bibinfo {author} {\bibfnamefont {U.}~\bibnamefont
  {Thiele}},\ }\bibfield  {title} {\bibinfo {title} {{Nonreciprocal
  Cahn-Hilliard Model Emerges as a Universal Amplitude Equation}},\ }\href
  {https://doi.org/10.1103/physrevlett.131.107201} {\bibfield  {journal}
  {\bibinfo  {journal} {Phys Rev Lett}\ }\textbf {\bibinfo {volume} {131}},\
  \bibinfo {pages} {107201} (\bibinfo {year} {2023})}\BibitemShut {NoStop}%
\bibitem [{\citenamefont {Matthews}\ and\ \citenamefont
  {Cox}(2000)}]{Matthews:2000}%
  \BibitemOpen
  \bibfield  {author} {\bibinfo {author} {\bibfnamefont {P.}~\bibnamefont
  {Matthews}}\ and\ \bibinfo {author} {\bibfnamefont {S.~M.}\ \bibnamefont
  {Cox}},\ }\bibfield  {title} {\bibinfo {title} {Pattern formation with a
  conservation law},\ }\href {https://doi.org/10.1088/0951-7715/13/4/317}
  {\bibfield  {journal} {\bibinfo  {journal} {Nonlinearity}\ }\textbf {\bibinfo
  {volume} {13}},\ \bibinfo {pages} {1293} (\bibinfo {year}
  {2000})}\BibitemShut {NoStop}%
\bibitem [{\citenamefont {Liu}\ and\ \citenamefont
  {Goldenfeld}(1989)}]{Liu:1989}%
  \BibitemOpen
  \bibfield  {author} {\bibinfo {author} {\bibfnamefont {F.}~\bibnamefont
  {Liu}}\ and\ \bibinfo {author} {\bibfnamefont {N.}~\bibnamefont
  {Goldenfeld}},\ }\bibfield  {title} {\bibinfo {title} {{Dynamics of phase
  separation in block copolymer melts}},\ }\href
  {https://doi.org/10.1103/physreva.39.4805} {\bibfield  {journal} {\bibinfo
  {journal} {Phys Rev A}\ }\textbf {\bibinfo {volume} {39}},\ \bibinfo {pages}
  {4805} (\bibinfo {year} {1989})}\BibitemShut {NoStop}%
\bibitem [{\citenamefont {Lipowsky}(2022)}]{Lipowsky:2022}%
  \BibitemOpen
  \bibfield  {author} {\bibinfo {author} {\bibfnamefont {R.}~\bibnamefont
  {Lipowsky}},\ }\bibfield  {title} {\bibinfo {title} {{Remodeling of Membrane
  Shape and Topology by Curvature Elasticity and Membrane Tension}},\ }\href
  {https://doi.org/10.1002/adbi.202101020} {\bibfield  {journal} {\bibinfo
  {journal} {Adv Biol}\ }\textbf {\bibinfo {volume} {6}},\ \bibinfo {pages}
  {2101020} (\bibinfo {year} {2022})}\BibitemShut {NoStop}%
\bibitem [{\citenamefont {Hassinger}\ \emph {et~al.}(2017)\citenamefont
  {Hassinger}, \citenamefont {Oster}, \citenamefont {Drubin},\ and\
  \citenamefont {Rangamani}}]{Hassinger:2017}%
  \BibitemOpen
  \bibfield  {author} {\bibinfo {author} {\bibfnamefont {J.}~\bibnamefont
  {Hassinger}}, \bibinfo {author} {\bibfnamefont {G.}~\bibnamefont {Oster}},
  \bibinfo {author} {\bibfnamefont {D.}~\bibnamefont {Drubin}},\ and\ \bibinfo
  {author} {\bibfnamefont {P.}~\bibnamefont {Rangamani}},\ }\bibfield  {title}
  {\bibinfo {title} {{Design principles for robust vesiculation in
  clathrin-mediated endocytosis}},\ }\href@noop {} {\bibfield  {journal}
  {\bibinfo  {journal} {Biophysical Journal}\ }\textbf {\bibinfo {volume}
  {112}},\ \bibinfo {pages} {310a} (\bibinfo {year} {2017})}\BibitemShut
  {NoStop}%
\bibitem [{\citenamefont {Sch\"uler}\ \emph {et~al.}(2014)\citenamefont
  {Sch\"uler}, \citenamefont {Alonso}, \citenamefont {Torcini},\ and\
  \citenamefont {B\"ar}}]{Schuler2014Spatio}%
  \BibitemOpen
  \bibfield  {author} {\bibinfo {author} {\bibfnamefont {D.}~\bibnamefont
  {Sch\"uler}}, \bibinfo {author} {\bibfnamefont {S.}~\bibnamefont {Alonso}},
  \bibinfo {author} {\bibfnamefont {A.}~\bibnamefont {Torcini}},\ and\ \bibinfo
  {author} {\bibfnamefont {M.}~\bibnamefont {B\"ar}},\ }\bibfield  {title}
  {\bibinfo {title} {{Spatio-temporal dynamics induced by competing
  instabilities in two asymmetrically coupled nonlinear evolution equations}},\
  }\href {https://doi.org/10.1063/1.4905017} {\bibfield  {journal} {\bibinfo
  {journal} {Chaos}\ }\textbf {\bibinfo {volume} {24}},\ \bibinfo {pages}
  {043142} (\bibinfo {year} {2014})}\BibitemShut {NoStop}%
\bibitem [{\citenamefont {Saha}\ \emph {et~al.}(2020)\citenamefont {Saha},
  \citenamefont {Agudo-Canalejo},\ and\ \citenamefont
  {Golestanian}}]{Saha2020Scalar}%
  \BibitemOpen
  \bibfield  {author} {\bibinfo {author} {\bibfnamefont {S.}~\bibnamefont
  {Saha}}, \bibinfo {author} {\bibfnamefont {J.}~\bibnamefont
  {Agudo-Canalejo}},\ and\ \bibinfo {author} {\bibfnamefont {R.}~\bibnamefont
  {Golestanian}},\ }\bibfield  {title} {\bibinfo {title} {{Scalar Active
  Mixtures: The Nonreciprocal Cahn-Hilliard Model}},\ }\href
  {https://doi.org/10.1103/physrevx.10.041009} {\bibfield  {journal} {\bibinfo
  {journal} {Phys Rev X}\ }\textbf {\bibinfo {volume} {10}},\ \bibinfo {pages}
  {041009} (\bibinfo {year} {2020})}\BibitemShut {NoStop}%
\bibitem [{\citenamefont {Brauns}\ and\ \citenamefont
  {Marchetti}(2024)}]{Brauns2024Nonreciprocal}%
  \BibitemOpen
  \bibfield  {author} {\bibinfo {author} {\bibfnamefont {F.}~\bibnamefont
  {Brauns}}\ and\ \bibinfo {author} {\bibfnamefont {M.~C.}\ \bibnamefont
  {Marchetti}},\ }\bibfield  {title} {\bibinfo {title} {{Nonreciprocal Pattern
  Formation of Conserved Fields}},\ }\href
  {https://doi.org/10.1103/physrevx.14.021014} {\bibfield  {journal} {\bibinfo
  {journal} {Phys Rev X}\ }\textbf {\bibinfo {volume} {14}},\ \bibinfo {pages}
  {021014} (\bibinfo {year} {2024})}\BibitemShut {NoStop}%
\bibitem [{\citenamefont {Radszuweit}\ \emph {et~al.}(2013)\citenamefont
  {Radszuweit}, \citenamefont {Alonso}, \citenamefont {Engel},\ and\
  \citenamefont {B\"ar}}]{Radszuweit2013Intracellular}%
  \BibitemOpen
  \bibfield  {author} {\bibinfo {author} {\bibfnamefont {M.}~\bibnamefont
  {Radszuweit}}, \bibinfo {author} {\bibfnamefont {S.}~\bibnamefont {Alonso}},
  \bibinfo {author} {\bibfnamefont {H.}~\bibnamefont {Engel}},\ and\ \bibinfo
  {author} {\bibfnamefont {M.}~\bibnamefont {B\"ar}},\ }\bibfield  {title}
  {\bibinfo {title} {{Intracellular Mechanochemical Waves in an Active
  Poroelastic Model}},\ }\href {https://doi.org/10.1103/physrevlett.110.138102}
  {\bibfield  {journal} {\bibinfo  {journal} {Phys Rev Lett}\ }\textbf
  {\bibinfo {volume} {110}},\ \bibinfo {pages} {138102} (\bibinfo {year}
  {2013})}\BibitemShut {NoStop}%
\bibitem [{\citenamefont {Edelstein-Keshet}\ \emph {et~al.}(2013)\citenamefont
  {Edelstein-Keshet}, \citenamefont {Holmes}, \citenamefont {Zajac},\ and\
  \citenamefont {Dutot}}]{Edelstein-Keshet2013From}%
  \BibitemOpen
  \bibfield  {author} {\bibinfo {author} {\bibfnamefont {L.}~\bibnamefont
  {Edelstein-Keshet}}, \bibinfo {author} {\bibfnamefont {W.~R.}\ \bibnamefont
  {Holmes}}, \bibinfo {author} {\bibfnamefont {M.}~\bibnamefont {Zajac}},\ and\
  \bibinfo {author} {\bibfnamefont {M.}~\bibnamefont {Dutot}},\ }\bibfield
  {title} {\bibinfo {title} {{From simple to detailed models for cell
  polarization}},\ }\href {https://doi.org/10.1098/rstb.2013.0003} {\bibfield
  {journal} {\bibinfo  {journal} {Phil T R Soc B}\ }\textbf {\bibinfo {volume}
  {368}},\ \bibinfo {pages} {20130003} (\bibinfo {year} {2013})}\BibitemShut
  {NoStop}%
\bibitem [{\citenamefont {Halatek}\ and\ \citenamefont
  {Frey}(2018)}]{Halatek:2018:Rethinking}%
  \BibitemOpen
  \bibfield  {author} {\bibinfo {author} {\bibfnamefont {J.}~\bibnamefont
  {Halatek}}\ and\ \bibinfo {author} {\bibfnamefont {E.}~\bibnamefont {Frey}},\
  }\bibfield  {title} {\bibinfo {title} {{Rethinking pattern formation in
  reaction–diffusion systems}},\ }\href
  {https://doi.org/10.1038/s41567-017-0040-5} {\bibfield  {journal} {\bibinfo
  {journal} {Nat Phys}\ }\textbf {\bibinfo {volume} {14}},\ \bibinfo {pages}
  {507} (\bibinfo {year} {2018})}\BibitemShut {NoStop}%
\bibitem [{\citenamefont {Brauns}\ \emph {et~al.}(2020)\citenamefont {Brauns},
  \citenamefont {Halatek},\ and\ \citenamefont {Frey}}]{Brauns2020Phase}%
  \BibitemOpen
  \bibfield  {author} {\bibinfo {author} {\bibfnamefont {F.}~\bibnamefont
  {Brauns}}, \bibinfo {author} {\bibfnamefont {J.}~\bibnamefont {Halatek}},\
  and\ \bibinfo {author} {\bibfnamefont {E.}~\bibnamefont {Frey}},\ }\bibfield
  {title} {\bibinfo {title} {{Phase-Space Geometry of Mass-Conserving
  Reaction-Diffusion Dynamics}},\ }\href
  {https://doi.org/10.1103/physrevx.10.041036} {\bibfield  {journal} {\bibinfo
  {journal} {Phys Rev X}\ }\textbf {\bibinfo {volume} {10}},\ \bibinfo {pages}
  {041036} (\bibinfo {year} {2020})}\BibitemShut {NoStop}%
\bibitem [{\citenamefont {Bergmann}\ and\ \citenamefont
  {Zimmermann}(2019)}]{Bergmann2019System}%
  \BibitemOpen
  \bibfield  {author} {\bibinfo {author} {\bibfnamefont {F.}~\bibnamefont
  {Bergmann}}\ and\ \bibinfo {author} {\bibfnamefont {W.}~\bibnamefont
  {Zimmermann}},\ }\bibfield  {title} {\bibinfo {title} {{On system-spanning
  demixing properties of cell polarization}},\ }\href
  {https://doi.org/10.1371/journal.pone.0218328} {\bibfield  {journal}
  {\bibinfo  {journal} {PLoS ONE}\ }\textbf {\bibinfo {volume} {14}},\ \bibinfo
  {pages} {e0218328} (\bibinfo {year} {2019})}\BibitemShut {NoStop}%
\bibitem [{\citenamefont {Brauns}\ \emph {et~al.}(2021)\citenamefont {Brauns},
  \citenamefont {Weyer}, \citenamefont {Halatek}, \citenamefont {Yoon},\ and\
  \citenamefont {Frey}}]{Brauns:2021}%
  \BibitemOpen
  \bibfield  {author} {\bibinfo {author} {\bibfnamefont {F.}~\bibnamefont
  {Brauns}}, \bibinfo {author} {\bibfnamefont {H.}~\bibnamefont {Weyer}},
  \bibinfo {author} {\bibfnamefont {J.}~\bibnamefont {Halatek}}, \bibinfo
  {author} {\bibfnamefont {J.}~\bibnamefont {Yoon}},\ and\ \bibinfo {author}
  {\bibfnamefont {E.}~\bibnamefont {Frey}},\ }\bibfield  {title} {\bibinfo
  {title} {{Wavelength Selection by Interrupted Coarsening in
  Reaction-Diffusion Systems}},\ }\href
  {https://doi.org/10.1103/physrevlett.126.104101} {\bibfield  {journal}
  {\bibinfo  {journal} {Phys Rev Lett}\ }\textbf {\bibinfo {volume} {126}},\
  \bibinfo {pages} {104101} (\bibinfo {year} {2021})}\BibitemShut {NoStop}%
\bibitem [{\citenamefont {{COMSOL AB}}(2023)}]{COMSOL}%
  \BibitemOpen
  \bibfield  {author} {\bibinfo {author} {\bibnamefont {{COMSOL AB}}},\
  }\href@noop {} {\bibinfo {title} {Comsol multiphysics {V}ersion 6.1}}
  (\bibinfo {year} {2023}),\ \bibinfo {note} {stockholm, Sweden}\BibitemShut
  {NoStop}%
\bibitem [{sup(2024)}]{supplemental_notebook}%
  \BibitemOpen
  \href@noop {} {\bibinfo {title} {{Supplemental Material for "Phase separation
  on Deformable Membranes: interplay of mechanical coupling and dynamic surface
  geometry"}}},\ \bibinfo {howpublished} {Available online} (\bibinfo {year}
  {2024}),\ \bibinfo {note} {see Supplemental Mathematica notebook, COMSOL file
  and videos at \url{link-provided-after-publication}.}\BibitemShut {Stop}%
\end{thebibliography}
\end{document}